%% file: SPHINCS_BSSN_v1.0_revised.tex
\documentclass[utf8]{FrontiersinHarvard} 

\setcitestyle{square} 
\usepackage{url,hyperref,lineno,microtype}
\usepackage[onehalfspacing]{setspace}

\input{preamble}

\def\keyFont{\fontsize{8}{11}\helveticabold }
\def\firstAuthorLast{Rosswog, Torsello \& Diener} 
\def\Authors{Stephan Rosswog\,$^{1,2}$, Francesco Torsello\,$^{2}$ and Peter Diener\,$^{3,4,*}$} 
\def\Address{$^{1}$University of Hamburg, Hamburger Sternwarte, Gojenbergsweg 112, 21029, Hamburg, Germany\\
$^{2}$The Oskar Klein Centre, Department of Astronomy, AlbaNova, Stockholm University, SE-106 91 Stockholm, Sweden\\
$^3$Center for Computation \& Technology, Louisiana State University, Baton Rouge, LA 70803, USA\\
$^4$ Department of Physics \& Astronomy, Louisiana State University, Baton Rouge, LA 70803, USA}
\def\corrAuthor{Peter Diener}
\def\corrEmail{diener@cct.lsu.edu}

\begin{document}
\onecolumn
\firstpage{1}

\def\xg{$\vec{r}^{\rm G}$}
\def\Xg{\vec{r}^{\rm G}}
\def\vr{\vec{r}}
\def\spB{\texttt{SPHINCS\_BSSN}}
\def\spB1{\texttt{SPHINCS\_BSSN\_v1.0}}
\def\Ma{\texttt{MAGMA2} }
\def\etk{\texttt{Einstein Toolkit}}
\def\kuibit{\texttt{kuibit}}
\def\p{\partial}
\def\msun{M$_{\odot}$}
\def\Msun{M$_{\odot}$ }
\def\be{\begin{equation}}
\def\ee{\end{equation}}
\def\bi{\begin{itemize}}
\def\i{\item}
\def\ei{\end{itemize}}
\def\ben{\begin{enumerate}}
\def\een{\end{enumerate}}
\def\bea{\begin{eqnarray}}
\def\eea{\end{eqnarray}}
\def\bt{\begin{tabbing}}
\def\et{\end{tabbing}}
\def\red{\textcolor{red}}
\def\blue{\textcolor{blue}}
\def\green{\textcolor{green}}
\def\ye{$Y_e$}
\def\Ye{$Y_e$ }
\def\rs{$R_{\rm S}$}
\def\Rs{$R_{\rm S}$ }
\def\noin{\noindent}
\def\QIone{\mathcal{Q}_{\; \; 1}}
\def\QItwo{\mathcal{Q}_{\; \; 2}}
\def\QIthree{\mathcal{Q}_{\; \; 3}}
\def\QIfour{\mathcal{Q}_{\; \; 4}}
\newcommand{\ud}{\mathrm{d}}
\newcommand{\sn}{\ensuremath{\mathrm{sn}}}
\newcommand{\cn}{\ensuremath{\mathrm{cn}}}
\newcommand{\dn}{\ensuremath{\mathrm{dn}}}
\newcommand{\sech}{\ensuremath{\mathrm{sech}}}
\newcommand{\nd}{\noindent}
\newcommand{\enth}{\mathcal{E}}
\newcommand{\carb}{$^{12}$C}
\newcommand{\Carb}{$^{12}$C $\;$}
\newcommand{\ox}{$^{16}$O}
\newcommand{\Ox}{$^{16}$O $\;$}
\newcommand{\tlg}{\tilde{\gamma}}
\newcommand{\tlG}{\tilde{\Gamma}}
\newcommand{\emfp}{e^{-4\phi}}
\newcommand{\tlA}{\tilde{A}}
\newcommand{\tlR}{\tilde{R}}
\newcommand{\xt}{\tilde{x}}
\newcommand{\dt}[1]{\partial_t #1}
\newcommand{\pdu}[3]{\partial_{#3} #1^{#2}}
\newcommand{\pdl}[3]{\partial_{#3} #1_{#2}}
\newcommand{\pdpdu}[4]{\partial_{#3}\partial_{#4} #1^{#2}}
\newcommand{\pdpdl}[4]{\partial_{#3}\partial_{#4} #1_{#2}}
\newcommand{\upwindu}[3]{\beta^{#3}\bar{\partial}_{#3} #1^{#2}}
\newcommand{\upwindl}[3]{\beta^{#3}\bar{\partial}_{#3} #1_{#2}}
\newcommand{\tlGn}{\tilde{\Gamma}_{\mathrm{(n)}}}
\newcommand{\tlGmixed}[2]{\tlG_{#1}^{\;\;\;#2}}
\newcommand{\apj}{Ap. J., }
\newcommand{\apjl}{Ap. J. Lett., }
\newcommand{\apjs}{Ap. J. Supp., }
\newcommand{\physrep}{Phys. Rep., }
\newcommand{\mnras}{Mon. not. RAS., }
\newcommand{\apss}{Astroph. \& Space Sci., }
\newcommand{\aap}{Astron. \& Astrophys., }
\newcommand{\prc}{Phys. Rev. C.,} 
\newcommand{\prd}{Phys. Rev. D.,} 
\newcommand{\nat}{Nature,}
\newcommand{\pasp}{Publications of the Astronomical Society of the Pacific,}

\title {The Lagrangian Numerical Relativity code SPHINCS\_BSSN\_v1.0} 

\author[\firstAuthorLast ]{\Authors} 
\address{} 
\correspondance{} 

\extraAuth{}

\maketitle

\begin{abstract}
We present version 1.0 of our Lagrangian Numerical Relativity code \texttt{SPHINCS\_BSSN}. This code evolves the full set of Einstein equations, but contrary to other Numerical Relativity codes, it evolves the matter fluid via 
Lagrangian particles in the framework of a high-accuracy version of Smooth Particle Hydrodynamics (SPH). The major new elements introduced here are: i) a new method to map the stress--energy tensor
(known at the particles) to the spacetime mesh, based on a local regression estimate; ii) additional measures that ensure the robust evolution of a neutron star through its collapse to a black hole; and iii) further refinements
in how we place the SPH particles for our initial data.
The latter are implemented in our code \spi which now, in addition to \lorene, can also couple to initial data produced by the initial data library \fuka. We 
discuss several simulations of neutron star mergers performed with \spB1, including 
irrotational cases with and without prompt collapse and a 
system where only one of the stars has a large spin ($\chi = 0.5)$.
\tiny
 \keyFont{ \section{Keywords:} Numerical Relativity, gravitational waves, neutron stars, Smooth Particle Hydrodynamics, Initial Data} 
\end{abstract}

\section{Introduction}

The first neutron star merger event, GW170817, opened the era of
gravitational wave-based multi-messenger astrophysics with  a 
bang. The inspiral stages were recorded in the detectors for about one 
minute  \citep{abbott17b}, and subsequently a firework was observed all across
the electromagnetic spectrum \citep{abbott17c}. Starting with a (special) short gamma-ray
burst (sGRB) detected $1.7~$s after the peak gravitational wave (GW) emission \citep{goldstein17,savchenko17}, followed by an
initially blue and subsequently rapidly reddening kilonova \citep{arcavi17,evans17,cowperthwaite17}. The event
also displayed a rising X-ray flux starting nine days after merger \citep{troja17}, peaking
after 160 days, that then started to decline steeply afterwards \citep{hallinan17,margutti17,lyman18,troja18}. This was interpreted as the imprints of a structured jet
observed at an angle of $\sim 25^\circ$ from the jet core \citep{lamb17,fong17,lamb18,hotokezaka18a}. Several years
later, broad-band synchroton afterglow was detected \citep{hajela22,troja22,hajela22}
that was interpreted as a ``kilonova afterglow" due to a mildly relativistic ejection
component interacting with the 
interstellar medium \citep{nakar11a,hotokezaka15a,hotokezaka18a}.

These observations allowed for a number of remarkable conclusions to be drawn.
For example, the pre-merger gravitational wave signal placed constraints on
the neutron star tidal deformability and therefore on the nuclear 
matter equation of state \citep{abbott17b}. The time delay between the 
GW peak and the sGRB showed that GWs propagate  
at the speed of light to an enormous precision \citep{abbott17d}. The event also allowed  for
a determination of the Hubble parameter \citep{abbott17a} completely independent of
previous approaches.
The bolometric light curve evolution of 
the kilonova was remarkably consistent with a broad range of 
decaying r-process elements \citep{metzger10b,kasen17,rosswog18a}, thereby 
confirming the long-held suspicion that neutron star mergers are
major sources of r-process elements in the cosmos \citep{lattimer74,symbalisty82,eichler89,rosswog99,freiburghaus99b}. 
While one would naively expect a red kilonova due to the extreme
neutron-richness of the original neutron star material and the related large opacities \citep{kasen13a}, 
the early blue kilonova emission underlined that about 0.01 \Msun
of the ejecta contained {\em light} (nucleon numbers $A <130$) r-process material, which is also
consistent with the identification of strontium lines \citep{watson19}. 
While underlining that a {\em broad} range of heavy elements has 
been produced, these observations also stress the importance 
of weak interactions that have transformed a substantial 
fraction of the neutrons into protons to produce the light r-process 
material. The ``kilonova afterglow" in turn, hints at a broad velocity 
distribution within the ejecta, extending to at least mildly relativistic 
velocities ($\gtrsim 0.6c$).

The phenomena described above illustrate the richness of the properties 
of the ejected material and they stress the importance of understanding the detailed properties of this 
$\sim 1\%$ of the total neutron star binary mass.  
Numerical simulations play an integral part in understanding and 
interpreting multi-messenger observations of compact binary systems.
The initial generations
of simulation models had a strong focus on the strong-field spacetime
dynamics, the motion of the neutron star fluid in it and the resulting 
gravitational wave emission, often with highly idealized microphysics.
Today's frontiers, however, have shifted more towards a complex multi-physics modelling in which General Relativity/strong-field gravity is only one ingredient out of  many. Apart from including
more physical processes such as magnetic field evolution or neutrino
transport, the now observationally established connection with the 
electromagnetic emission also places higher demands on the length and time 
scales that need to be modelled in a neutron star merger event.

The vast majority of today's Numerical Relativity codes makes use 
of Eulerian hydrodynamics. While these codes have produced a multitude 
of important results, a larger methodological variety would be 
desirable, for independent checks of results but also to potentially 
address problems where established methods struggle, as, for example, 
in the long-term evolution of merger ejecta. The \texttt{SPHINCS\_BSSN}
code is the first Lagrangian hydrodynamics code that solves the 
full set of Einstein equations. The first results, restricted to 
standard relativistic 
hydrodynamics tests and oscillating and collapsing neutron stars, were 
presented in \cite{rosswog21a}. In \cite{diener22a} the scope was 
extended to neutron star mergers, at that stage using simple polytropic 
equations of state and \lorene -based initial conditions \citep{Grandclement_2001,Gourgoulhon_2001,lorene}
produced with an extension of the ``Artificial Pressure Method", 
originally proposed in \cite{rosswog20a}, to the case of neutron star 
binaries. Subsequently,  further technical improvements were introduced 
\citep{rosswog22b} and nuclear matter properties were approximated in 
terms of piecewise polytropic equations of state \citep{read09}.
We have further improved our simulation technology and in this paper, we 
describe the ingredients of what we have tagged as ``version 1.0" of our 
code, \spB1. We emphasize the latest improvements while still giving
a broad overview of the complete methodology. The new elements include a further refined method
to map the particle properties (specifically their stress--energy tensor)
to the spacetime mesh and additional measures to ensure that we
can robustly evolve a neutron star through its collapse to a black hole. 
We also describe further refinements in the particle setup of our initial data, produced with the code \spi \cite{sphincsid}. 

Our article is structured as follows. In Section~\ref{sec:method}
we describe the methodology, with Section~\ref{sec:hydro} focusing on the hydrodynamics, Section~\ref{sec:EOS} on the equation of state, Section~\ref{sec:SpT} on the spacetime evolution
and Section~\ref{sec:M_ST_coupling} on how spacetime and matter evolution
are coupled. In Section~\ref{sec:stable_BH_formation} we describe measures to robustly
evolve the collapse of a neutron star to a black hole and in 
Section~\ref{sec:id} we summarize our latest improvements in constructing
SPH initial conditions in full General Relativity. In Section~\ref{sec:results}
we discuss several examples of neutron star 
mergers, and we summarize our results in
Section~\ref{sec:summary}

\section{Methodology of SPHINCS\_BSSN\_v1.0}
\label{sec:method}
Here we describe the methodological elements of the  
code \spB1. We only concisely summarize 
those parts that have been laid out already elsewhere in the literature, 
but we describe in detail the  elements that are presented here for the first time. These 
are in particular: a) a more sophisticated coupling between the spacetime 
and the fluid (via a local polynomial regression estimate),
b) specific measures (enhancement of grid resolution and the potential transformation of
fluid into ``dust" particles) that enable us to robustly simulate the formation of black holes.

\subsection{Hydrodynamics}
\label{sec:hydro}
The hydrodynamic evolution equations in \SpB are modelled via a high-accuracy 
version of Smooth Particle Hydrodynamics (SPH), see \cite{monaghan05,rosswog09b,springel10a,price12a} and \cite{rosswog15c} for reviews of the method.
The basics of the relativistic SPH equations have been derived very explicitly in Section 4.2 of \cite{rosswog09b} 
and we will only present the final equations here. Several accuracy-enhancing elements such as kernels, 
gradient estimators and dissipation steering strategies (for either Newtonian or relativistic cases) have been explored 
in a recent series of papers  \cite{rosswog15b,rosswog20a,rosswog20b} and most of them are also implemented in \spB1.\\
We use units in which $G=c=1$ and masses are measured in solar units. These ``code units" approximately correspond in physical units to $1.47663~$km for lengths, to  $4.92549 \times 10^{-6}$ s for
time and to $6.17797 \times 10^{17}$ gcm$^{-3}$ for densities. We further
use the metric signature (-,+,+,+) and we measure all energies in units of $m_0 c^2$, where $m_0$ is the average baryon mass\footnote{This quantity depends on the actual nuclear composition, but simply using the atomic mass unit $m_u$ gives
a precision of better than 1\%. We therefore use  the approximation $m_0 \approx m_u$ in the following. A more detailed discussion can be found in Section 2.1.1 of \cite{diener22a}}. 
Greek indices run from  0 to 3 and latin indices from 1 to 3. Contravariant spatial indices of a vector quantity $\vec{w}$ at particle $a$ are 
denoted as $w_a^i$, while covariant ones will be written as $(w_i)_a$.\\
To discretize our fluid equations we choose a ``computing frame" in which the computations are performed. Quantities in this frame usually differ
from those calculated in the local fluid rest frame. 
The line element in a 3+1 split of spacetime reads (e.g. \cite{baumgarte10})
\be
ds^2= -\alpha^2 dt^2 + \gamma_{ij} (dx^i + \beta^i dt) (dx^j + \beta^j dt),
\ee
where $\alpha$ is the lapse function, $\beta^i$ the shift vector and $\gamma_{ij}$ the spatial 3-metric.
We use a generalized Lorentz factor 
\be
\Theta\equiv \frac{1}{\sqrt{-g_{\mu\nu} v^\mu v^\nu}} \quad {\rm with} \quad v^\alpha=\frac{dx^\alpha}{dt}.
\label{eq:theta_def}
\ee
The coordinate velocities $v^\alpha$ are related to the four-velocities, normalized as $U_\alpha U^\alpha=-1$, by
\be
v^\alpha= \frac{dx^\alpha}{dt}= \frac{U^\alpha}{\Theta}= \frac{U^\alpha}{U^0}.
\label{eq:v_mu}
\ee
We choose the computing frame baryon number density $N$ as density variable, which is related to the
baryon number density as measured in the local fluid rest frame, $n$, by 
\be
N= \sqrt{-g}\, \Theta\, n.
\label{eq:N_def}
\ee
Here $g$ is the determinant of the spacetime metric.  Note that this density variable is  very similar to 
what is used in Eulerian approaches \citep{alcubierre08,baumgarte10,rezzolla13a,shibata16}.
We keep the baryon number of each SPH particle, $\nu_a$, constant so that exact numerical baryon 
number conservation is hard-wired. At every (Runge--Kutta  sub-)step, the computing frame baryon number density
at the position of a particle $a$ is  calculated via a weighted sum (actually  very similar to 
Newtonian SPH)
\be
N_a= \sum_b \nu_b\, W(|\vec{r_a} - \vec{r}_b|,h_a),
\label{eq:N_sum}
\ee
where the smoothing length $h_a$ characterizes the support size 
of the SPH smoothing kernel $W$, see below. As numerical momentum variable, 
we choose the canonical momentum per baryon 
\be
(S_i)_a = (\Theta \mathcal{E} v_i)_a,
\label{eq:can_mom}
\ee
where $\mathcal{E}= 1 + u + P/n$ is the relativistic enthalpy per baryon with $u$ being the internal energy per
baryon and $P$ the gas pressure. The quantity $S_i$ evolves according to
\be
\frac{d(S_i)_a}{dt}  =  \left(\frac{d(S_i)_a}{dt}\right)_{\rm hyd} +  \left(\frac{d(S_i)_a}{dt}\right)_{\rm met},
\label{eq:dSdt_full}
\ee
where the hydrodynamic part is given by
\be
\left(\frac{d(S_i)_a}{dt}\right)_{\rm hyd}= -\sum_b \nu_b \left\{ \frac{P_a}{N_a^2}  D^a_i  +  
\frac{P_b}{N_b^2} D^b_i \right\}
\label{eq:dSdt_hydro}
\ee
and the gravitational part by
\be 
\left(\frac{d(S_i)_a}{dt}\right)_{\rm met}= \left(\frac{\sqrt{-g}}{2N} T^{\mu \nu} \frac{\p g_{\mu \nu}}{\p x^i}\right)_a
\label{eq:dSdt_metric}.
\ee
Here we have used the abbreviations
\be
D^a_i \equiv   \sqrt{-g_a} \;  \frac{\p W_{ab}(h_a)}{\p x_a^i} \quad {\rm and} \quad 
D^b_i \equiv    \sqrt{-g_b} \; \frac{\p W_{ab}(h_b)}{\p x_a^i}
\label{eq:kernel_grad}
\ee
and $W_{ab}(h_k)$ is a shorthand for $W(|\vec{r}_a - \vec{r}_b|/h_k)$.
As numerical energy variable we use the canonical energy per baryon 
\be
e_a= \left(S_i v^i + \frac{1 + u}{\Theta}\right)_a = \left(\Theta \mathcal{E} v_i v^i + \frac{1 + u}{\Theta}\right)_a,
\label{eq:can_en}
\ee
which is evolved according to
\be
\frac{d e_a}{dt}= \left(\frac{d e_a}{dt}\right)_{\rm hyd}  + \left(\frac{de_a}{dt}\right)_{\rm met},
\label{eq:energy_equation}
\ee
with
\be
\left(\frac{d e_a}{dt}\right)_{\rm hyd} = -\sum_b \nu_b \left\{ \frac{P_a}{N_a^2}  \;  v_b^i   \; D^a_i +  
\frac{P_b}{N_b^2} \;  v_a^i \; D^b_i \right\}
\label{eq:dedt_hydro}
\ee
and
\be
\left(\frac{de_a}{dt}\right)_{\rm met}= -\left(\frac{\sqrt{-g}}{2N} T^{\mu \nu} \frac{\p g_{\mu \nu}}{\p t}\right)_a.
\label{eq:dedt_metric}
\ee

\begin{figure}
   \centering
   \includegraphics[width=0.7\columnwidth]{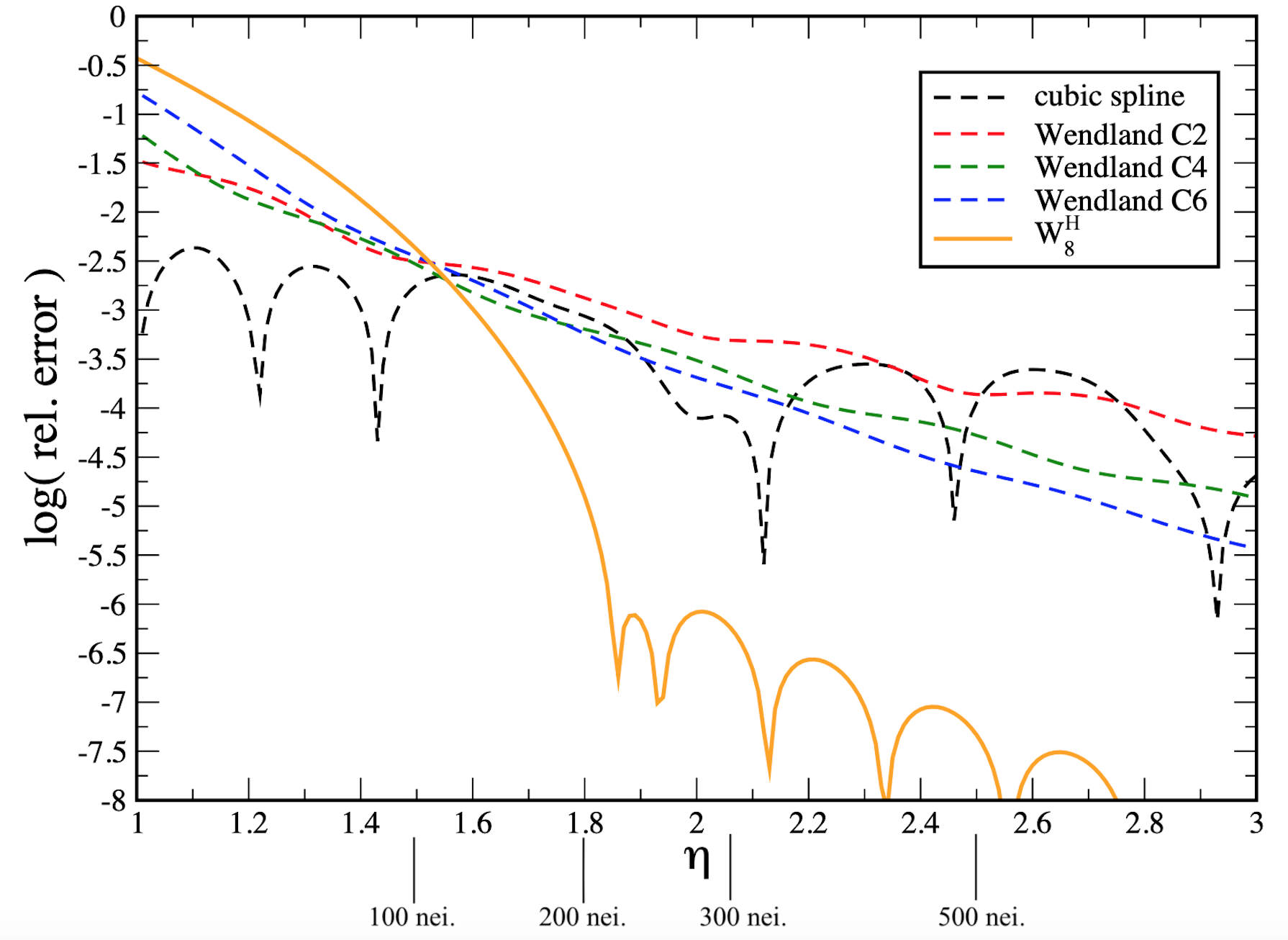} 
   \caption{Error on the density estimate for different kernels as a function of support size for particles arranged 
   on a cubic lattice. The parameter $\eta$ determines the smoothing length $h$ as a multiple of the typical particle
   spacing via $h= \eta (\nu/N)^{1/3}$, the support radius of all kernels is $2h$. We have also indicated the corresponding 
   number of neighbour particles.}
   \label{fig:kernel_accuracy}
\end{figure}

It is instructive to make the connection between our numerical momentum variable $S_i$, Equation~(\ref{eq:can_mom}), and the Arnowitt--Deser--Misner (ADM) momentum of the fluid. 
Given a spatial vector field $\xi^i$ which tends to a Cartesian basis vector at spatial infinity, the ADM linear momentum along the direction of $\xi^i$ is defined as \cite[Equation (8.40)]{gourgoulhon20123+1} (see also \cite[Section~II]{krishnan_quasi-local_2007})
\begin{align}
\label{eq:admmom}
    P_\xi^{\rm ADM}\coloneqq \dfrac{1}{8\pi}\int_{\partial\Sigma}\dd \sigma_j\left(K^j{}_i - \delta^j{}_i \,K\right)\xi^i,
\end{align}
where $\partial\Sigma$ is the boundary of a spacelike hypersurface $\Sigma$ that extends up to spatial infinity, $\dd \sigma_j$ its surface element, $K^j{}_i$ is the extrinsic curvature and $K=K^i{}_i$ its trace. Using Gauss' theorem, the momentum constraint, and some geometry, the integral in Equation~\eqref{eq:admmom} can be written as (see \autoref{app:admmomsph})
\begin{align}
\label{eq:admmomcons}
    P_\xi^{\rm ADM}&= \int_{\Sigma}\dd ^3x\sqrt{\gamma}
    \left[\xi^i\eulS_i + \dfrac{1}{16\pi} \left(K^{ij} - \gamma^{ij} \,K\right)\mathscr{L}_\xi\gamma_{ij}\right],
\end{align}
where $\gamma_{ij}$ is the spatial metric, $\eulS_i$ is the spatial part of the momentum density measured by the Eulerian observer $\eulS_\rho\coloneqq -n^\mu\,T_{\mu\nu}\,\gamma^\nu{}_\rho$, and $\mathscr{L}_\xi$ is the Lie derivative along $\xi^i$. The two terms in the integrand of Equation~(\eqref{eq:admmomcons}) are the parts of the ADM linear momentum determined by the fluid and the spacetime, respectively.  The expression in 
Equation~(\eqref{eq:admmomcons}) allows us to write the ADM momentum \emph{of the fluid} in terms of the SPH canonical momentum $\sphS_i$, after we relate the latter with the Eulerian spatial momentum density $\eulS_i$. We do this explicitly in \autoref{app:admmomsph}. Here we show the final result and its SPH approximation:
\begin{align}
\label{eq:admmomesph}
    P_\xi^{\rm ADM,fluid}&= \int_{\Sigma}\dd ^3x\sqrt{\gamma}\,\xi^i \eulS_i=\int_{\Sigma}\dd ^3x \,N \,\xi^i \sphS_i 
    \simeq \sum_b \nu_b \left(\xi^i \sphS_i \right)_b
\end{align}
with the index $b$ running over all the particles, $\nu_b$ being the baryon number of particle $b$, $N$ defined in Equation~(\ref{eq:N_def}), $\eulS_i=\Theta\, n\,\alpha\,\sphS_i$, $\alpha$ is the lapse function, and $\sqrt{-g}=\alpha\sqrt{\gamma}$ \cite[Equation (2.124)]{baumgarte10}. The rightmost formula in Equation~(\ref{eq:admmomesph}) can be used to compute an estimate of the ADM momentum of the fluid using only SPH fields. Such an estimate can then be compared with another one computed using Equation~(\ref{eq:admmom}); this comparison tells us about the error on the ADM momentum of the fluid introduced when we model the fluid with SPH particles. We make this comparison at the level of the initial data (ID) in Section~\ref{subsec:admlinmom}.\\
For the  kernel function that is needed for the SPH approximation, we have implemented a large variety
of choices. We have performed many numerical 
experiments similar to the one shown in Figure~\ref{fig:kernel_accuracy}, some of which
are documented in detail in \cite{rosswog15b}. Here we 
only provide the details of our preferred kernels. The first of these favorites is
the Wendland C6-smooth kernel \citep{wendland95}
\be
W(q)= W^{WC6} (q)= \frac{\sigma}{h^3} (1-q)^8_+ (32 q^3 + 25 q^2 + 8 q + 1),
\ee
where we use exactly 300 constributing neighbour particles. The normalization constant is $\sigma= 1365/(64 \pi)$ in 3D and the symbol $(.)_+$ denotes the cutoff function max$(.,0)$. Other favorites include (some) 
members of the family  of the harmonic-like kernels (\cite{cabezon08})  
\begin{align}
            \hspace{-5mm} W (q) = W^{\rm H}_n(q)= \frac{\sigma_n}{h^3} \left\{\begin{array}{llr}
        1, & \; q=0,\\[4mm]
        \left( \frac{\sin\left[ \frac{\pi}{2} q \right]}{\frac{\pi}{2} q}\right)^n
          & \; 0< q <2,\\[4mm]
        0, & \; {\rm else},
       \end{array}\right. 
       \label{eq:WH8_kernel}
\end{align}
namely those with $n=7$ and 8.
Out of this family, we chose, after ample experiments, the kernel with
$n=8$ for which we use exactly 220 contributing neighbour particles. For
this kernel the normalization constant is $\sigma_8= 1.17851074088357$ 
in 3D. Contrary to Wendland kernels \citep{wendland95}, this kernel is 
{\em not} immune against the (benign) pairing instability,
but it provides an excellent density estimate. We show in Figure~\ref{fig:kernel_accuracy} the result of a density measurement 
experiment. Particles are placed on a cubic lattice and masses are assigned so that the mass density 
is exactly unity. We then use several kernel functions, the commonly used cubic spline kernel \citep{monaghan92},
the Wendland C2, C4, and C6 kernels (see \cite{dehnen12} for the explicit expressions) and the $W^{\rm H}_8$ kernel,
Equation~(\ref{eq:WH8_kernel}).
The smoothing lengths $h$ in this experiment are set as multiples of the typical particle separation $(\nu/N)^{1/3}$ ,
$h= \eta (\nu/N)^{1/3}$, and each of the kernels has a support radius of $2h$. We have also indicated some approximate
numbers of contributing particles (``neighbours")  for our cubic lattice. For neighbour numbers just beyond 200, the density 
accuracy in this experiment is about three orders of magnitude better for $W^{\rm H}_8$ compared to the Wendland C6 kernel.
While it is a priori not entirely clear how to weigh the excellent performance in this idealized experiment against the desirable
property of the Wendland kernels to maintain a very regular particle distribution during dynamical evolution \citep{rosswog15b},
we choose in this work the $W^{\rm H}_8$-kernel and we find very satisfactory results. On inspecting the simulations 
presented here,  we do not see any significant pairing among the particles and the density distributions appear noise-free
while exhibiting sharp features. \\
To keep the numerical noise at a minimum, we choose at each time step the 
smoothing length of each particle $a$, $h_a$, so that there are exactly  220 contributing neighbour particles within 
the support radius $2h_a$. In other words, at
each time step the smoothing length of particle $a$, $h_a$, is 
set so that $2 h_a$ equals  the distance to the 221-th closest SPH particle. This particle sits by definition at the radius
where the kernel becomes zero, so
that exactly 220 particles have a non-zero contribution. This approach ensures a very smooth and subtle evolution of the smoothing length and avoids the introduction of noise through the update of the smoothing length.
In practice, this is achieved by using our fast RCB-tree \citep{gafton11},
for more details on the exact  procedure, we refer to the \Ma code paper \citep{rosswog20a} where the same approach was used. A large number of experiments confirms that we get very similar
results when using $W^{WC6}$ and $W^{H}_8$.\\
To robustly handle relativistic shocks, our momentum and energy equations are augmented with dissipative
terms. These terms consist of an artificial dissipation and an artificial conductivity part. In both of these terms
we apply a slope-limited reconstruction to the mid-point of each particle pair and this reconstruction approach has been shown
to massively reduce unwanted dissipative effects \citep{rosswog20a}. In order to further reduce dissipation
where it is not needed, we make our dissipation parameters time-dependent; they are increased when a shock or
numerical noise is detected, but otherwise they decay exponentially to a small value (here chosen as $\alpha_0= 0.1$). 
Since no changes compared to our previous work have been made, we refer the interested reader for the equations, 
implementation details and tests to \cite{rosswog22b}.\\
The quantities that we evolve numerically, $S_i$ and $e$, together with the density $N$, see Equation(\ref{eq:N_sum}),
are numerically very convenient, but they are not the physical quantities that we are interested in. 
We therefore have to recover the physical quantities $n$, $u$, $v^i$, $P$ from  
$N$, $S_i$, $e$ at every integration (sub-)step. This ``recovery'' step is done in a very similar way as in Eulerian approaches.
For polytropic equations of state, our strategy is described in detail in Section~2.2.4 of \cite{rosswog21a},
and the modifications needed for piecewise polytropic EOSs are laid out explicitly in Appendix A of \cite{rosswog22b}. 

\subsection{Equations of state (EOS)}
\label{sec:EOS}
To close the system of hydrodynamic equations we need an equation of state. Currently, we are using 
piecewise polytropic approximations to cold nuclear matter equations of state \citep{read09}, that 
are enhanced with an ideal gas-type thermal contribution to both pressure and specific internal energy, a 
common approach in Numerical Relativity simulations. For explicit expressions please see Appendix A of \cite{rosswog22b}. 
To date, we have implemented 14 piecewise polytropic equations of state, but since the effects
coming from different EOSs are not the topic of this
code paper, we restrict ourselves to results
obtained for the APR3 EOS \citep{akmal98} only. This EOS allows for a maximum mass of $M_{\rm TOV}^{\rm max}= 2.39$ \Msun and a 1.4 \Msun neutron star has a dimensionless tidal deformability of $\Lambda_{1.4}= 390$. Indirect arguments and the statistics of the radio pulsar/X-ray neutron star distribution point to  values in the range of $\sim 2.2-2.4$  \Msun \citep{fryer15,antoniadis16,margalit17,bauswein17,shibata17c,rezzolla18,alsing18} for the ``best educated guess" of the maximum neutron star mass
and a recent Bayesian study \citep{biswas22}  suggests a maximum TOV mass of 2.52$^{+0.33}_{-0.29}$ \msun, all broadly consistent with our choice of the APR3 EOS.
  More sophisticated treatments of high-density nuclear matter physics will be addressed in the future.

\subsection{Spacetime evolution}
\label{sec:SpT}
We evolve the spacetime according to the (``$\Phi$-version'' of the) BSSN
equations \citep{shibata95,baumgarte99}. We have written wrappers around code extracted
from the well tested \McL thorn of the Einstein 
Toolkit \cite{loeffler12,ETK:web}. The dynamical variables used in this method are related to the
Arnowitt--Deser--Misner (ADM) variables $\gamma_{ij}$ (3-metric), $K_{ij}$ (extrinsic curvature),
$\alpha$ (lapse function) and $\beta^{i}$ (shift vector) and they read
\begin{align}
  \phi & =  \frac{1}{12} \log(\gamma), \\
  \tlg_{ij} & =  \emfp \gamma_{ij}, \\
  K & =  \gamma^{ij} K_{ij}, \\
  \tlG^{i} & =  \tlg^{jk} \tlG^{i}_{jk}, \\
  \tlA_{ij} & =  \emfp\left ( K_{ij}-\frac{1}{3}\gamma_{ij} K\right ),
\end{align}
where $\gamma = \operatorname{det}(\gamma_{ij})$,  $\tlG^{i}_{jk}$ are the
Christoffel symbols related to the conformal metric $\tlg_{ij}$ and $ \tlA_{ij}$ is
the conformally rescaled, traceless part of the extrinsic curvature. The corresponding
evolution equations read
\begin{align}
  \dt{\phi} & = -\frac{1}{6} \left ( \alpha K - \pdu{\beta}{i}{i} \right) + \upwindu{\phi}{}{i}, \label{eq:BSSN1}\\
 \dt{\tlg_{ij}} & = -2\alpha \tlA_{ij} + \tlg_{ik} \pdu{\beta}{k}{j}
                   + \tlg_{jk} \pdu{\beta}{k}{i}
                    -\frac{2}{3} \tlg_{ij} \pdu{\beta}{k}{k}
                         + \upwindl{\tlg}{ij}{k}, \\
  \dt{K} & = -\emfp \left ( \tlg^{ij} \left [ \pdpdu{\alpha}{}{i}{j}
               +2\pdu{\phi}{}{i}\pdu{\alpha}{}{j} \right ]
               - \tlGn^{i}\pdu{\alpha}{}{i} \right ) 
         + \alpha \left ( \tlA^{i}_{j} \tlA^{j}_{i} +\frac{1}{3} K^2
               \right ) + \upwindu{K}{}{i} + 4 \pi \alpha ( \rho + s ), \\
  \dt{\tlG^{i}} & = -2 \tlA^{ij} \pdu{\alpha}{}{j} + 2 \alpha \left (
                    \tlG^{i}_{jk} \tlA^{jk} - \frac{2}{3} \tlg^{ij}
                    \pdu{K}{}{j} +
                  6 \tlA^{ij} \pdu{\phi}{}{j}\right )
                 \nonumber\\
                  &                     +\tlg^{jk} \pdpdu{\beta}{i}{j}{k} + \frac{1}{3}
                    \tlg^{ij} \pdpdu{\beta}{k}{j}{k} -\tlGn^{j}\pdu{\beta}{i}{j}
                    + \frac{2}{3} \tlGn^{i}\pdu{\beta}{j}{j} 
                    + \upwindu{\tlG}{i}{j} -16 \pi \alpha \tlg^{ij} s_j, \\
  \dt{\tlA_{ij}} & = \emfp \left [ -\pdpdu{\alpha}{}{i}{j} + \tlG^{k}_{ij}
                       \pdu{\alpha}{}{k} + 2 \left ( \pdu{\alpha}{}{i}
                       \pdu{\phi}{}{j}+\pdu{\alpha}{}{j} \pdu{\phi}{}{i}\right )
                       +\alpha R_{ij} \right ]^{\mathrm{TF}} +\alpha ( K \tlA_{ij}- 2 \tlA_{ik} \tlA^{k}_{j} ) \nonumber\\
                    & + \tlA_{ik} \pdu{\beta}{k}{j}
                       + \tlA_{jk} \pdu{\beta}{k}{i}
                       - \frac{2}{3} \tlA_{ij} \pdu{\beta}{k}{k} 
                 +\upwindl{\tlA}{ij}{k} - \emfp \alpha 8 \pi
                       \left (T_{ij}-\frac{1}{3} \gamma_{ij} s\right ),
\end{align}
where
\begin{align}
  \rho & =  \frac{1}{\alpha^2} ( T_{00} - 2 \beta^{i} T_{0i} +
             \beta^{i}\beta^{j} T_{ij} ),\label{eq:BSSN_rho} \\
  s & =  \gamma^{ij} T_{ij}, \\
  s_{i} & =  -\frac{1}{\alpha} ( T_{0i} - \beta^{j} T_{ij}),\label{eq:BSSN_Si}
\end{align}
and $\upwindu{}{}{i}$ denote partial derivatives that are upwinded based on the
shift vector. The superscript ``TF" in the evolution equation of $\tlA_{ij}$ denotes 
the trace-free part of the bracketed term.
Finally $R_{ij} = \tlR_{ij} + R^{\phi}_{ij}$, where
\begin{align}
  \tlG_{ijk} & =  \frac{1}{2}\left ( \pdl{\tlg}{ij}{k} + \pdl{\tlg}{ik}{j}
               - \pdl{\tlg}{jk}{i} \right ), \\
  \tlGmixed{ij}{k} & =  \tlg^{kl} \tlG_{ijl}, \\
  \tlG^{i}_{jk} & =  \tlg^{il}\tlG_{ljk}, \\
  \tlGn^{i} & =  \tlg^{jk} \tlG^{i}_{jk}, \\
  \tlR_{ij} & =  -\frac{1}{2} \tlg^{kl} \pdpdl{\tlg}{ij}{k}{l}
                  +\tlg_{k(i} \pdu{\tlG}{k}{j)}
                  +\tlGn^{k} \tlG_{(ij)k} 
            +\tlG^{k}_{il} \tlGmixed{jk}{l}
                  +\tlG^{k}_{jl} \tlGmixed{ik}{l}
                  +\tlG^{k}_{il} \tlGmixed{kj}{l}, \\
  R^{\phi}_{ij} & =  -2\left (\pdpdu{\phi}{}{i}{j}
                 -\tlG^{k}_{ij}\pdu{\phi}{}{k}\right )
                 -2\tlg_{ij} \tlg^{kl} 
            \left ( \pdpdu{\phi}{}{k}{l}
                 -\tlG^{m}_{kl}\pdu{\phi}{}{m}\right )
                + 4\pdu{\phi}{}{i}\pdu{\phi}{}{j}
             - 4\tlg_{ij}\tlg^{kl}\pdu{\phi}{}{k}\pdu{\phi}{}{l}.
\end{align}
The derivatives on the right-hand side of the BSSN equations are evaluated via standard Finite Differencing
techniques and, unless mentioned otherwise, we use sixth order differencing as a default. We have recently implemented
a fixed mesh refinement for the spacetime evolution, which is described in detail in \cite{diener22a}, to which
we refer the interested reader. For the gauge choices, we use a variant of ``1+log-slicing," where the lapse
is evolved according to
\be
\partial_t \alpha= -2 \alpha K
\ee
and a variant of the ``$\Gamma$-driver" shift condition with
\be
\partial_t \beta^i= \frac{3}{4}(\tilde{\Gamma}^i-\eta \beta^i),
\ee
where $\eta$ is the ``$\beta$-driver'' parameter.

\subsection{Coupling between spacetime and matter}
\label{sec:M_ST_coupling}
The \SpB approach of evolving the spacetime on a mesh and the matter fluid via particles
requires a continuous information exchange: the gravity part of the particle
evolution is driven by derivatives of the metric, see Eqs.~(\ref{eq:dSdt_metric}) and (\ref{eq:dedt_metric}),
which are known on the mesh, while the stress--energy tensor, needed in Equation~(\ref{eq:BSSN1})-(\ref{eq:BSSN_Si}), 
is known on the particles. This bidirectional information exchange is needed at every Runge--Kutta substep;
in our case, with an optimal 3$^{\rm rd}$ order Runge--Kutta algorithm \citep{gottlieb98}, three times per numerical time step.\\
The mesh-to-particle mapping step is performed via 5$^{\rm th}$ order Hermite interpolation, that we developed \citep{rosswog21a} 
extending the work of \cite{timmes00a}. Contrary to a standard Lagrange polynomial interpolation, the Hermite 
interpolation guarantees that the metric remains twice differentiable as particles pass from one grid cell to another
and therefore avoids the introduction of additional  noise. Our approach is explained in detail in Section 2.4 of 
\cite{rosswog21a} to which we refer the interested reader.\\
Here, we provide a further improvement to the more challenging of the two steps, the particle-to-mesh mapping. As in our previous work \citep{diener22a,rosswog22b}, we follow a ``Multi-dimensional Optimal Order Detection" (MOOD) strategy. The mapping is performed simultaneously with different orders, and the most accurate
solution is selected out of the possible results (according to some error measure, see below), provided that the solution meets 
additional physical admissibility conditions.\\
This is somewhat similar in spirit to 
ENO reconstruction schemes in mesh-based hydrodynamics,
see \cite{zhang16} for a concise summary of ENO and WENO schemes, 
in the sense that one does not pretend to know a priori which
reconstruction order (or here: interpolation order) is "good enough". Instead one tries a variety of options and selects
the best one. Our scheme is similar to ENO methods where one tries different stencils and selects the smoothest one according to a suitable criterion. (WENO actually goes one step further by using a weighted sum of the different options to combine them into potentially higher order.)\\
Whereas  in \cite{rosswog22b} we used kernel functions of different orders borrowed from
vortex methods \citep{cottet00,cottet14}, we determine here the best fit to the 
particle properties at a grid point by using the local polynomial regression estimate that we describe in the next section.

\subsubsection{Local Regression Estimate (LRE)}
\begin{figure}
   \centering
   \includegraphics[width=\columnwidth]{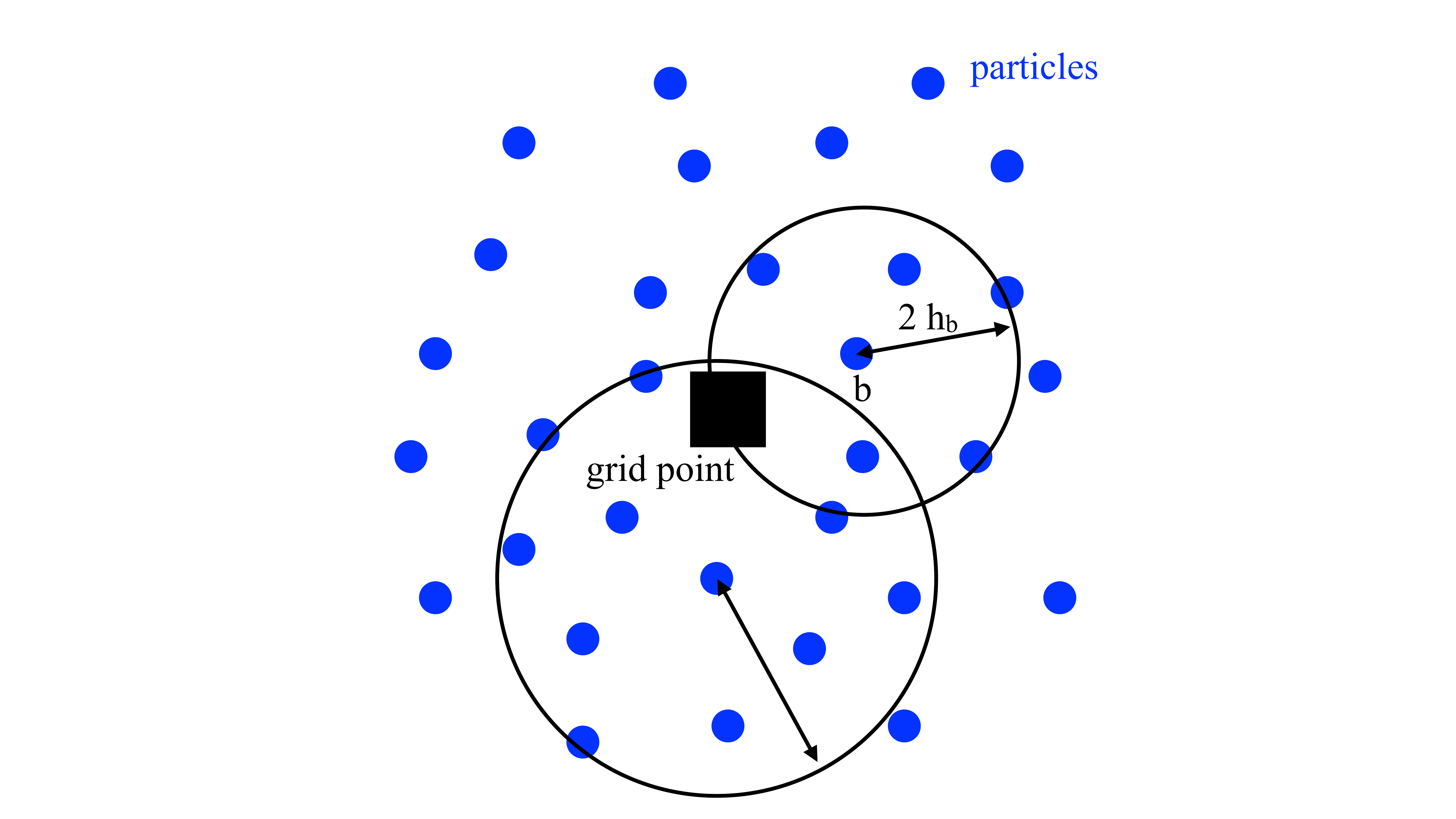} 
   \caption{Geometry of the particle-to-mesh mapping: function values that
   are known at particle positions (blue circles) are to be mapped to a grid point (black square).}
   \label{fig:sketch}
\end{figure}

The task is to find the function that best fits the values of the particles surrounding the grid point $\vec{r}^{\rm G}$, see Figure~\ref{fig:sketch}.
If we assume that the values at the particle positions $\{ f^p\}$ are described by a function $f(\vec{r})$,
we can Taylor-expand this function around the desired grid point $\vec{r}^{\rm G}$:
\bea
f(\vec{r})=&& f(\vec{r}^{\rm G}) + (\p_i f)_{\Xg} (\vec{r} - \Xg)^i + \frac{1}{2!} (\p_{ij} f)_{\Xg} \; (\vec{r} - \Xg)^i  (\vec{r} - \Xg)^j \nonumber \\
                &+& \frac{1}{3!} (\p_{ijk} f)_{\Xg} \; (\vec{r} - \Xg)^i  (\vec{r} - \Xg)^j  (\vec{r} - \Xg)^k \nonumber \\
                &+& \frac{1}{4!} (\p_{ijkl}  f)_{\Xg} \; (\vec{r} - \Xg)^i  (\vec{r} - \Xg)^j  (\vec{r} - \Xg)^k  (\vec{r} - \Xg)^l  \nonumber\\
                &+& \mbox{higher order terms.}
\eea
This Taylor expansion can be interpreted as a polynomial approximation of a given order where the basis functions have been
shifted to the point of interest \xg. The local approximation $\tilde{f}(\vec{r})$ of $f(\vec{r})$ around the point \xg then reads
\be
\tilde{f}^G(\vr)= \vec{\beta}^G \cdot \vec{P}^G(\vr),
\label{eq:func_approx}
\ee 
where the coefficient vector reads
  \begin{align}
   \vec{\beta}^G = \Big[& f(\vec{r}^{\rm G}),\;\;
           (\p_x f)_{\Xg},\,
           (\p_y f)_{\Xg},\,
           (\p_z f)_{\Xg},\;\;\nonumber \\
           &
           \frac{1}{2}(\p_{xx} f)_{\Xg},\,
           (\p_{xy} f)_{\Xg},\,
           (\p_{xz} f)_{\Xg},\,
           \frac{1}{2}(\p_{yy} f)_{\Xg},\,
           (\p_{yz} f)_{\Xg},\,
           \frac{1}{2}(\p_{zz} f)_{\Xg},\;\;
           \cdots\Big],\label{eq:coeff}
\end{align}
and the ``shifted basis functions" read
   \begin{align}
   \vec{P}^G(\vec{r}) = \Big [ &
           1, \;\;
           \Delta^G x,\,
           \Delta^G y,\,
           \Delta^G z, \;\; \\
           & \Delta^G x \Delta^G x,\,
           \Delta^G x \Delta^G y,\,
           \Delta^G x \Delta^G z,\,
           \Delta^G y \Delta^G y,\,
           \Delta^G y \Delta^G z, \,
           \Delta^G z \Delta^G z,\;\;
        \cdots\Big]^T,
   \end{align}
  where $\Delta^G \vec{r}=\vec{r}-\vec{r}^{\rm G}= [\Delta^G x, \Delta^G y,\Delta^G z]^T$,
  and $\Delta^G x= x - x^G$ (similarly for the other components).
The degrees of freedom ($\mathrm{DoF}$) for a given number of dimensions $d$ and a maximum polynomial order of the basis $m$ are given by
\be
\mathrm{DoF}= \frac{(d+m)!}{d!\,m!},
\ee
that is, in 3D, we have 1, 4, 10, 20 and 35 DoF for constant, linear, quadratic, cubic and quartic polynomials.\\
The optimal coefficients at $\vec{r}^G$, $\vec{\beta}^G$, are found by
minimizing the error functional
\be
\epsilon^G \equiv  \sum_p [f_p - \tilde{f}^G(\vec{r}_p)]^2 \; W_{pG} = \sum_p \left[f_p - { \sum_{i=1}^{\mathrm{DoF}} \beta^G_i P^G_i(\vec{r}_p)} \right]^2 \; W_{pG},
\ee
where $W_{pG}= W(\vec{r}^p-\vec{r}^G,l_p)$ is a smooth weighting function that ensures that particles further away 
from the grid point are weighted less in the error functional, and the $p$-sum runs over all contributing particles. The kernel width is set by $l_p= (\nu_p/N_p)^{1/3}$.
One could choose, for example, compactly supported SPH kernels for this weighting function; we choose simple tensor products of 1D $M_4$-kernels. The exact form of the weight function does not have a strong 
influence on the resulting error measure.\\
The optimal coefficients that minimize the error at $\vec{r}^G$ are determined via
\be
\left(\frac{\p \epsilon^G}{\p \beta_i^G }\right)_{\Xg} \stackrel{!}{=} 0
\ee
and, with a few steps of algebra, 
one finds
\be
\beta^G_i = \left(M_{ik}\right)^{-1} \; B_k,
\label{eq:coeff2}
\ee
where
\be
M_{ik}=  \sum_p P^G_i(\vec{r}_p) \; P^G_k(\vec{r}_p) \; W_{pG}
\ee
is the ``moment matrix" and
\be
B_k = \sum_p f_p \; P^G_k(\vec{r}_p) \; W_{pG}
\ee
is a vector depending on the function values at the particles. Since the moment matrix
does not depend on the function values themselves (only on the relative positions),
it can be used for several function vectors (here one for each $T_{\mu\nu}$ component).
With increasing polynomial order the condition number of the moment matrix $\mathcal{C}\equiv ||M M^{-1}||$, a measure for how close a matrix is to being singular, can become
very large, therefore we use the singular value decomposition \citep{press92}
to solve the system in Equation~(\ref{eq:coeff2}). The condition numbers, however, can be massively reduced by using re-scaled basis functions which we illustrate in Appendix \ref{sec:LRE_appendix} and Figure~\ref{fig:condition_number}. In practice, we use re-scaled basis functions.\\
The function value estimate at the grid point (see Equation~(\ref{eq:coeff})),
$\tilde{f}(\vec{r}^G)$ is the first component of $\beta^G$, the derivatives are $\p_x\tilde{f}(\vec{r}^G), \p_y\tilde{f}(\vec{r}^G)$, 
$\p_z\tilde{f}(\vec{r}^G)$ are the components two to four, and so on. For the case where we only allow the lowest polynomial
order (i.e., a constant polynomial), the moment matrix has only one element
\be
M= \sum_p W_{pG} \Longrightarrow M^{-1}= \frac{1}{\sum_p W_{pG}},
\ee
and the function vector becomes
\be
B= \sum_p f_p W_{pG}.
\ee
In this case the function value at the grid point is estimated as
\be
\tilde{f}^G= \frac{\sum_p f_p W_{pG}}{\sum_p W_{pG}}.
\ee
In other words, this is just the straightforward kernel-weighted average of the 
values at the contributing particles (with an exact partition of unity enforced).
We show an example of the LRE function approximation in Appendix~\ref{sec:LRE_appendix}.

\subsubsection{An LRE-based MOOD approach}
While in a well-sampled region, as for example the stellar interior, a high-order LRE approximation
likely provides the most accurate function estimate, this is not necessarily true near the stellar surface. There, due to the Gibbs phenomenon, spurious oscillations may occur. Therefore, we calculate $T_{\mu\nu}$ 
estimates for different polynomial orders $m$, $\tilde{T}_{\mu\nu}^{{\rm G}, m}$, and then select the ``optimal order" 
that best represents the particle values and is physically admissible. We use the following error measure:
\bea
E^{{\rm G}, m} \equiv \sum_p W_{pG} \left[
\sum_{\mu,\nu} \left\{ {\tilde{T}}_{\mu \nu}^{{\rm G}, m} (\vr_p) - T_{\mu \nu,p} \right\}^2 \right]
=\sum_p W_{pG} \left[ \sum_{\mu,\nu} \left\{  \left( \vec{\beta}^{G, m}_{\mu,\nu} \cdot P^G(\vr_p) \right)    - T_{\mu \nu,p}   \right\}^2\right].
\label{eq:map_error}
\eea
In other words, based on the optimal coefficients at the grid point, 
we estimate the function values {\em at each particle position} that contributes and 
calculate the weighted quadratic deviation as error measure. 
This approach differs from our earlier one \citep{rosswog22b} by
not using pre-defined kernel functions, but instead applying the LRE-approximation, and by considering {\em all} $T_{\mu\nu}$ components
in the error measure, rather than just $T_{00}$ as before.\\
The most straightforward MOOD estimate
would be to calculate estimates for different polynomial orders $m$ and to select the solution with the smallest value
of $E^{{\rm G}, m}$. Unfortunately, this only works in {\em nearly} all cases. In very few cases, where a grid point lies
{\em outside} the neutron star surface, but the finite-size SPH particles still contribute to this point, we found that 
the approximation with the smallest error may deliver values much larger than those at the contributing particles,
or even unphysical values such as a negative $T_{00}$ (energy density). 
For these reasons, additional measures need to be taken near the stellar surface. However, the question to be answered is: how does an SPH particle know that it is near the surface? 

\subsubsection{Detecting not well-embedded grid points}
Here we use a simple, yet, as it turns out, very robust method to detect whether a grid point is well engulfed by particles
or not. In a first step, at each particle position we numerically calculate an estimate for an expression that has an
analytically known result and where the deviation from the exact result can be used to identify surface particles. 
For this purpose we chose $\nabla \cdot \vec{r}=3$ and the numerical estimate given by
\be
\left( \nabla \cdot \vec{r} \right)_a= \sum_b \frac{\nu_b}{N_b} (\vr_b - \vr_a) \cdot \nabla_a W_{ab}(h_a).
\label{eq:divx}
 \ee
This expression is one of the standard SPH discretizations (similar to the commonly used expression for $\nabla \cdot \vec{v}$,
see Equation~(31) in \cite{rosswog09b}). 
To avoid another neighbour-loop over all particles, expression (\ref{eq:divx}) can be conveniently calculated alongside the SPH derivatives. {This means that the update of $\nabla \cdot \vec{r}$  is lagging behind
by one third of a time step. The property of being at the surface changes on a much longer time scale and only
averages of the deviations are used, see below, so that using a 
value of $\nabla \cdot \vec{r}$ calculated a third of a time
step earlier is well suited for our purposes.
In deriving SPH expressions, surface terms are usually neglected and therefore
 the expression Equation~(\ref{eq:divx}) only yields an accurate approximation to the exact value of 3 if it is embedded
from all sides with particles. If instead the expression is evaluated near a surface, contributing particles are missing on one
side and therefore the numerical estimate is substantially smaller than the theoretical value. 
From the relative error 
\be
\delta_a \equiv \frac{|(\nabla \cdot \vec{r})_a - 3 |}{3}
\ee
we calculate the average deviation $\langle \delta \rangle_G= (\sum_{b=1}^{n}\delta_b)/n$ over the particles that contribute to the error measure Equation~(\ref{eq:map_error}).
If $\langle \delta \rangle_G$ is above a  given threshold, the corresponding grid point is identified as being outside the 
particle surface. 
We show an example from the inspiral of two neutron stars in Figure~\ref{fig:surface_id}. In the stellar interior, the values of $\langle \delta \rangle_G$ 
are $\approx 0.005$, while those at the surface reach $\sim 0.1$. After some experimentation, we have chosen a threshold of 0.05 for 
$\langle \delta \rangle_G$; for grid points at which $\langle \delta \rangle_G$ exceed this threshold, we only use the lowest order 
mapping, $m=0$, see below. This approach robustly avoids all "outlier points" in the mapping, and the mapped values of $T_{\mu\nu}$
accurately reflect the matter distribution. 
\begin{figure}
    \centering
    \includegraphics[width=0.9\columnwidth]{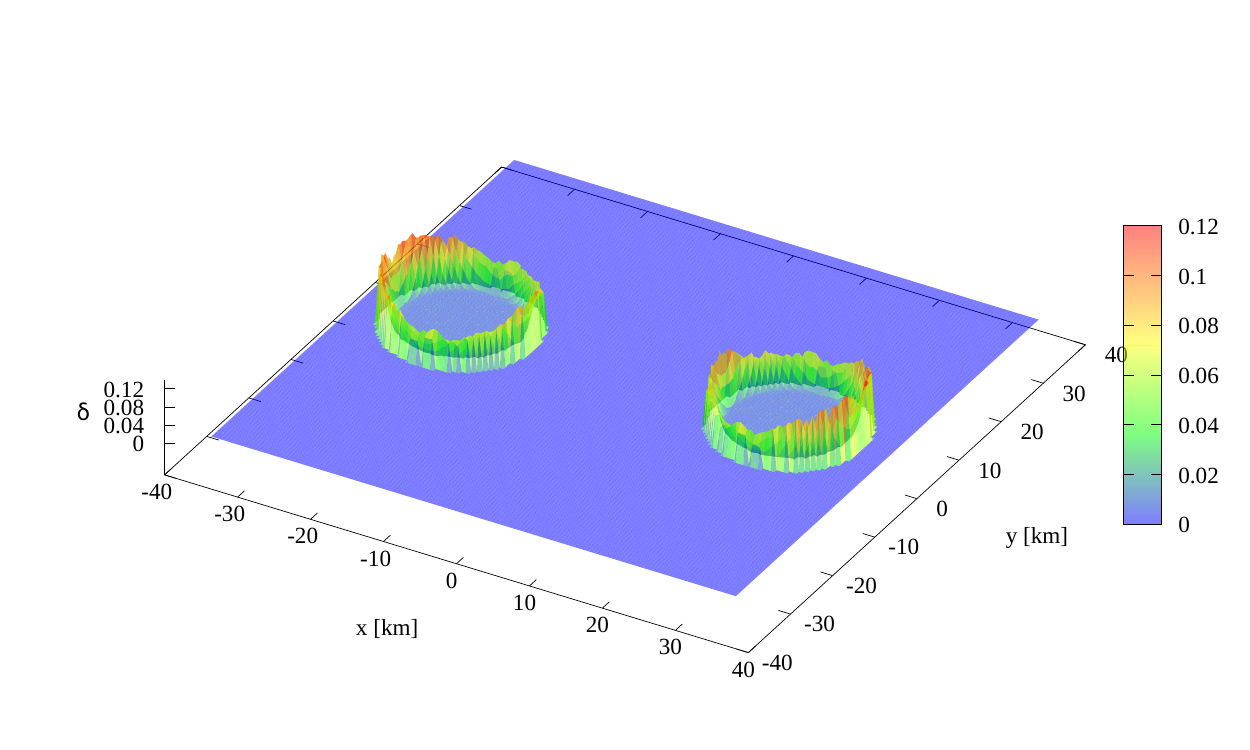}
    \caption{The average value of $\langle \delta \rangle_G$ for all the particles that
             contribute to the mapping of the stress--energy tensor at
             grid points in the $xy$-plane. For grid points inside the
             stars, the value is typically $\delta\approx 0.005$.}
    \label{fig:surface_id}
\end{figure}

\begin{figure}
   \centering
   \includegraphics[width=0.9\columnwidth]{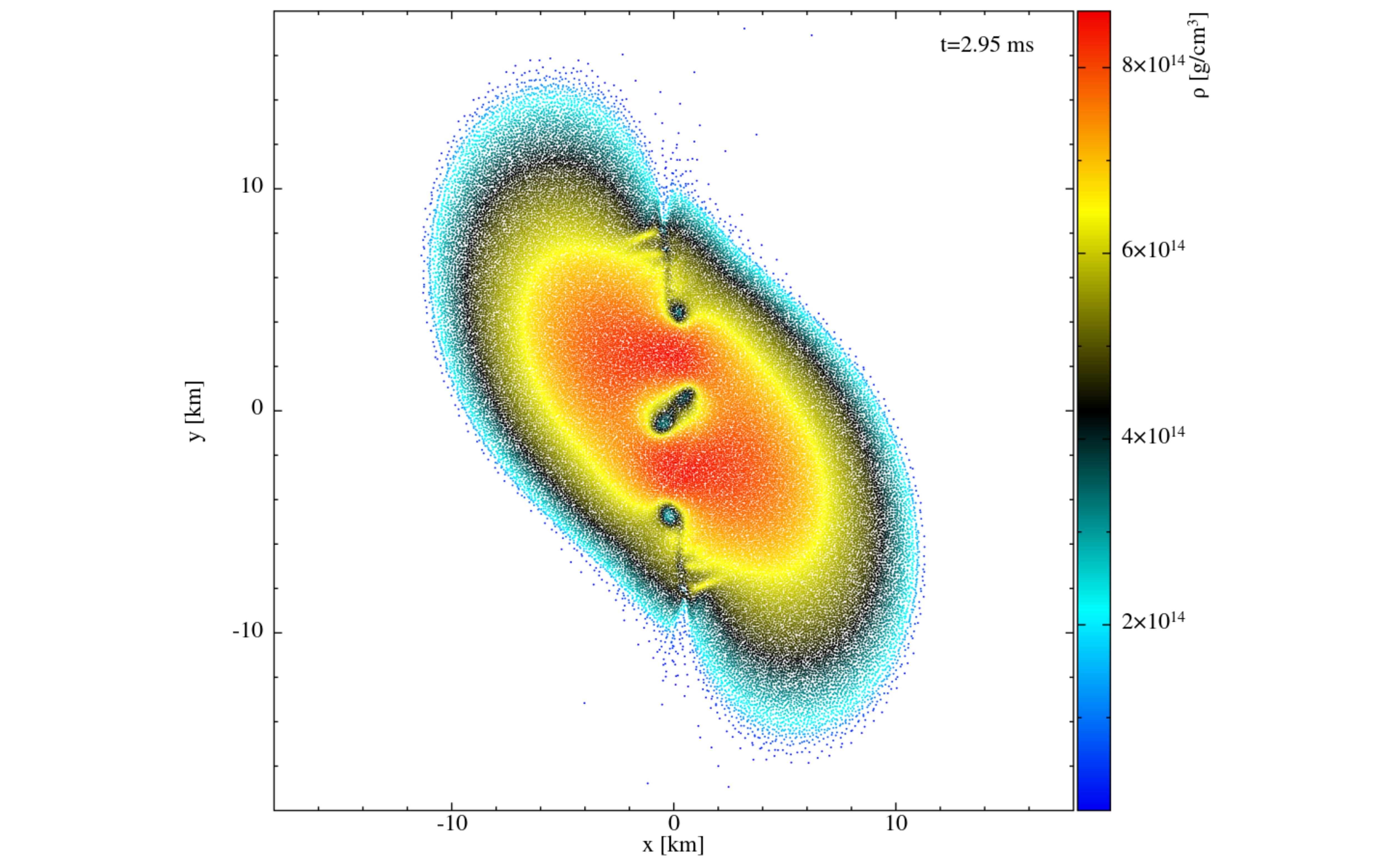}\vspace{-2cm}
   \hspace*{-1.6cm}\includegraphics[width=1.1\columnwidth]{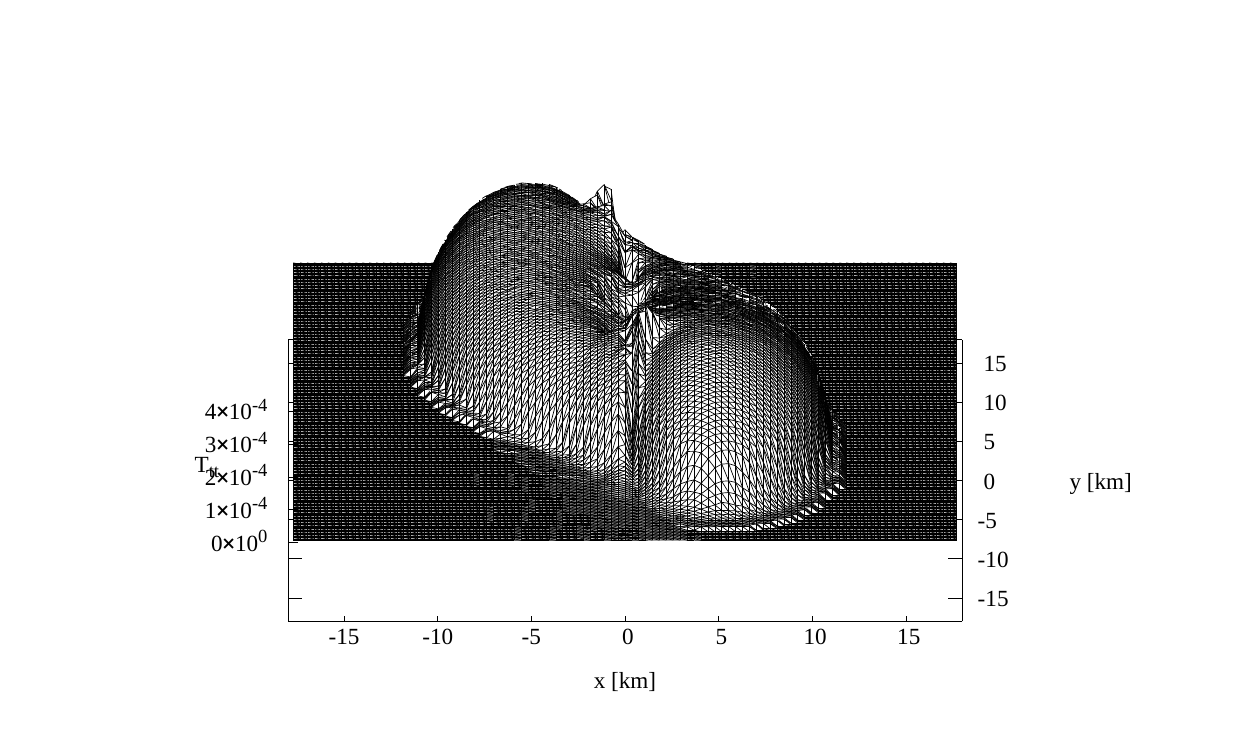}

    \vspace{-1cm}
    
   \caption{At the top, we show a cut of the density at $t=2.95 \mathrm{ms}$ for the case of a equal mass binary system with 2 1.5 \msun neutron stars with the APR3 equation of state. At the bottom, we show the corresponding mapping of the $tt$-component of the stress--energy tensor onto the grid.}
   \label{fig:tmunu_map}
\end{figure}

\subsubsection{Summary of the particle-to-mesh mapping algorithm}
The first step in the algorithm consists in identifying the particles that contribute to a given grid point $G$. This list of particles is identified via a hash-grid as described in 
detail in Section 2.1.3 of \cite{diener22a}. Subsequently, we perform the following steps:
\bi
\i calculate the LRE estimates $\tilde{T}_{\mu\nu}^{{\rm G}, m}$ for the polynomial orders $m= 0, 1, 2, 3$ using Equation~(\ref{eq:func_approx})
\i if $\langle \delta \rangle_G > 0.05$ choose $\tilde{T}_{\mu\nu}^{{\rm G}, m=0}$ since this is a grid point outside the particle surface
\i if more than 40 particles (= twice the number of degrees of freedom for cubic polynomials) contribute {\em and} the error
    $E^{G, 3}$ is smallest {\em and} $\tilde{T}_{00}^{{\rm G}, m=3} > 0$, choose $\tilde{T}_{\mu\nu}^{{\rm G}, m=3}$
\i if more than 20 particles (= twice the number of degrees of freedom for quadratic polynomials) contribute {\em and} the error
    $E^{G, 2}$ is smallest {\em and} $\tilde{T}_{00}^{{\rm G}, m=2} > 0$, choose $\tilde{T}_{\mu\nu}^{{\rm G}, m=2}$
\i if more than 8 particles (= twice the number of degrees of freedom for linear polynomials) contribute {\em and} the error
    $E^{G, 1}$ is smallest {\em and} $\tilde{T}_{00}^{{\rm G}, m=1} > 0$,  choose $\tilde{T}_{\mu\nu}^{{\rm G}, m=1}$
\i in all remaining cases choose the robust ``parachute method", that is, polynomial order 0 and $\tilde{T}_{\mu\nu}^{{\rm G}, m=0}$.        
\ei
The additional conditions on the number of contributing particles have been introduced to avoid
the inversion of poorly conditioned matrices. In Figure~\ref{fig:tmunu_map} we illustrate how well this
procedure works. The top plot shows a cut of the density of particles
at t = 2.95 ms for the case of an equal mass binary system with two
1.5 \Msun neutron stars with the APR3 equation of state. The bottom
plot shows the resulting mapped $tt$-component of the stress energy
tensor in the $xy$-plane of the grid. As can be seen, all the features, that are visible in the particle density profile, are clearly reproduced in the mapped stress-energy tensor.

\subsection{Stably simulating the formation of black holes}
\label{sec:stable_BH_formation}
The remnant of a binary NS merger can, depending on the EOS, the 
total mass and spin (and further processes which are not modelled here), 
undergo a collapse to a black hole (BH). This can happen either ``promptly" 
on the dynamical timescale of the remnant (typically $\sim 1~$ms),
or it can be ``delayed" for several dynamical timescales or, for binaries
at the low-mass end, it may not occur at all. If a BH does form, extra care
is required in order to avoid numerical problems. \\
The first potential problem we noticed when we initially attempted to simulate a 
collapse was simply that at some point in time, the particles
become packed so close 
together, that the grid resolution is insufficient to evolve the metric accurately
enough to maintain a physically valid solution. This can result in failures in the
recovery of the primitive variables.\\
To cure this problem, we allow for the addition of more refinement levels. After some
experiments, we decided to use the ratio of the hydrodynamic and the
BSSN time step as an indicator of when it is time  to add another refinement 
level.  As the particles move closer together, their Courant time step,
$\Delta t_{\rm SPH}= \xi_{\rm SPH} (h/c_s)$, decreases (here $c_s$ is the speed of sound),
whereas the Courant time step for the mesh,
$\Delta t_{\rm BSSN}= \xi_{\rm G} (\Delta x/c)$, stays constant. Note that we have
omitted particle and grid labels for readability, and for 
clarity we have written the expression using the speed of light $c$ explicitly. $\Delta x$ is the resolution on the finest grid.
For the dimensionless pre-factors, our default values are $\xi_{\rm SPH}= 0.2$
and $\xi_{\rm G}=0.35$. Whenever the ratio of the two time steps grows
beyond a  threshold, 
\begin{equation}
\frac{\Delta t_{\mathrm{BSSN}}}{\Delta t_{\mathrm{SPH}}}>C_{\mathrm{refine}}
\end{equation}
we add a refinement level. In numerical experiments, we found that $C_{\mathrm{refine}}=2.5$ works well.\\
Our current mesh refinement hierarchy is very simple: our finer grids
are always half the size of the next coarser grid, and they are centered at the
origin. We follow this strategy also when we create a new refinement level:
the new level has the same number of points and half the size of the previously finest grid, and it is
again centered at the origin. The data on the new grid are calculated from 
the data on the previously finest grid using the same interpolation operators 
that we use in the prolongation operators to fill the ghost cells of the refinement 
levels, i.e.\ via cubic Lagrange interpolation.\\
After adding a new refinement level, $\Delta t_{\mathrm{BSSN}}$ is
half its previous value and the time step ratio will again be less than
$C_{\mathrm{refine}}$. As the collapse proceeds, $\Delta t_{\mathrm{SPH}}$
will continue to decrease and may eventually trigger the addition of 
another refinement level. This can in principle continue indefinitely, but
would eventually slow the simulation to a halt due to very small time steps.
Therefore, at some point in time, we start to  remove particles in the innermost, collapsing core.\\
The best criterion would of course be to remove particles when they are deep enough
inside the forming black hole. Since we have not yet implemented an
apparent horizon finder, we do not know precisely when and where the black
hole forms at run time. Instead, we rely on the value of the lapse, $\alpha$, at the
location of the particles as a proxy. With the slicing and shift conditions used in
the code, it is observed that the value of $\alpha$ at the horizon of a
single static black hole, evolved to numerical stationarity, is around 0.3,
see e.g. Figs. 14 and 16 in \cite{rosswog21a}.
The value of $\alpha$ at the actual horizon during the dynamical phase
of the collapse will of course vary slightly, but will not differ too much 
from the value of 0.3. Thus, we remove particles when they enter regions 
that have a {\em substantially lower lapse} than this threshold value.\\
Based on the turduckening idea of 
\cite{Brown:2008sb}, we should be able to safely change the interior of
a black hole as long as it is done sufficiently deep inside. In particular
removing the source (the particles) of the stress energy tensor, should
not affect how the continuing collapse is seen from the outside. To be
safe we want to wait as long as possible before starting to remove 
particles, but as the particles pile up inside the black hole the
Courant time step decreases dramatically, potentially making the 
collapse process very computationally expensive. However, we are saved
by the observation that particles  are
essentially in free fall  when they are that deep inside a black hole. 
Hence, the energy momentum tensor is completely
dominated by the rest mass and velocity with the pressure and internal 
energy only providing negligible small corrections. Therefore we can
convert particles into ``dust'' by setting their pressure and internal
energy to zero once the lapse at their position is less than
$\alpha_{\mathrm{dust}}=0.05$.\\
The dust particles will no longer
affect the evolution of their neighbours and will no longer contribute
to $\Delta t_{\mathrm{SPH}}$. They will simply evolve along geodesics
until the lapse at their positions falls below 
$\alpha_{\mathrm{cut}}=0.02$ at which point we simply remove them
completely from the simulation. With this two stage process, converting
particles first to dust and then later removing them, we manage to have
the particles contribute to the stress energy tensor for significantly
longer without adversely affecting the time step. We found that removing
particles as soon as the lapse dropped below $\alpha=0.05$ could lead to
a delay in the collapse of the lapse in the center of the black hole.\\
Post-processing our data using the apparent horizon finder from the 
Einstein Toolkit \cite{loeffler12}, showed that this delay in the lapse
did not significantly affect the horizon properties, but we still prefer
to avoid it.\\
It turns out, that even with extra refinement levels, once we start
converting particles to dust and later removing
particles, it is still possible to eventually get failures in the recovery
of primitive variables. However, this only happens when the particles are
essentially in free fall and the solution is to convert them to dust
before they reach $\alpha_{\mathrm{dust}}$. This is only necessary in
very few cases, if at all.\\
Once particles have been converted to dust, we still have to recover
the primitive variables from the evolved variables, but this is very
simple.
The relations between the evolved and primitive variables, Eqs.(\ref{eq:N_def}), (\ref{eq:can_mom}) and (\ref{eq:can_en}), for a dust particle
with vanishing $u$ and $P$ reduce to (omitting for simplicity the particle label)
\begin{align}
  N & = \sqrt{-g}\,\Theta \, n \\
  S_i & = \Theta\, v_i \\
  e & = \Theta\, v_i v^i + \frac{1}{\Theta}.
\end{align}
That is, $S_i$ reduces to the spatial part of the covariant 4-velocity,
$U_{\mu}$.  Therefore, we can find $U_0$ so that the 4-velocity is properly
normalized, $U_{\mu}U^{\mu}=-1$. Raising the index on the covariant
4-velocity, we can simply read off $\Theta = U^0$ and then find $n$. 
In case a fluid particle is transformed to dust, we also adjust $e$ so that it is consistent with
the values of $v^i$ and $\Theta$. In summary, the recovery
of primitive from evolved variables is always possible for dust particles in a straightforward way, 
and the particles can keep evolving, contributing to the stress--energy tensor, 
but do not affect any of their neighbour
particles in any negative way, until they 
can be safely removed once 
their lapse value has dropped below the removal threshold.

\section{Improvements to the initial data setup}
\label{sec:id}

\input{sphincs_id}

\section{Numerical Results for Neutron star mergers}
\label{sec:results}
Here we show astrophysical examples of neutron star mergers with \spB1. Standard hydrodynamics
tests such as shock tube tests are not impacted by any of the new elements introduced here, therefore,
we refer the interested reader to our previous papers \citep{rosswog21a,rosswog22b}.
In Section~\ref{sec:stable_remnant} we show a binary neutron star merger where a remnant survives (for
at least several dynamical time scales), Section~\ref{sec:BH_formation} shows an example where the
merger remnant promptly collapses to form a black hole 
and in Section~\ref{sec:single_Spin} we show results
for a binary system where only one of the neutron
stars has a large spin, whereas the other has none. All
systems are of equal mass, the simulations start from an initial separation of 45 km and are performed with slightly more than 2 million SPH particles (except for the case where collapse to a black hole happens promptly; here 1 million SPH particles are used), the APR3 EOS,
initially 7 mesh refinement levels out to $\approx 2268$ km in each coordinate direction and a minimum
initial grid spacing of $\Delta x= 369$ m. Keep in mind that
new refinement levels are added dynamically, when the 
criterion described in Section~\ref{sec:stable_BH_formation} is met.
In the run presented in Section~\ref{sec:BH_formation} where there is a collapse to a black hole, 
the number of refinement levels dynamically increases
up to 11.

\subsection{Neutron star merger with surviving remnant}
\label{sec:stable_remnant}

We show here the merger of two 1.3 \Msun neutron stars, with initial conditions
produced by \lorene. After a few orbits of inspiral, the stars merge and remain initially close to perfect symmetry, see panel 1 in Figure~\ref{fig:dens_APR3_2x1.3}. The strong shear at the interface between the stars is Kelvin--Helmholtz unstable and as matter from this region is sprayed out, deviations from perfect symmetries emerge (panel 2),
as also frequently seen in Eulerian neutron star merger simulations. A few milliseconds later, the remnant settles into what seems a stationary state
with a bar-like central object shedding mass via a spiral-wave into the surrounding torus. This spiral wave ejection channel might have played an important
role in the early blue kilonova signal after the first observed neutron star merger GW170817 \citep{nedora19}, see \cite{rosswog22c} for a review.\\
In the left panel of Figure~\ref{fig:lapse_APR3_2x1.3} we show the
evolution of the maximum density
(red curve, right axis) together with 
the minimum lapse function value (black curve, left axis). In an initial very deep compression  the density reaches
a value close to $9.4 \times 10^{14}$ g cm$^{-3}$, then the remnant bounces back and, after several more oscillations, the peak density settles
near a value of $9.5 \times 10^{14}$ g cm$^{-3}$.
As expected, the lapse is lowest where the density is highest and vice versa. 
The right panel shows the value of the maximum GW amplitude times the distance to the observer, as calculated via the quadrupole approximation (orange) and as extracted from the spacetime via the Weyl scalar $\psi_4$ (black), how these are calculated in detail can be found in Appendix A of \cite{diener22a}. All $\Psi_4$ based GW waveforms were analyzed using \kuibit~\cite{kuibit}. Again, we find a rather good agreement of the quadrupole result with
the more sophisticated $\psi_4$ method. 
The radiated energy (red curve, left axis) and angular momentum (black curve, right axis) are plotted as a
percentage of the initial ADM values in the left panel of Fig~\ref{fig:dEdJ}. More than 2\% of the initial ADM mass and more than 20\% of the initial ADM angular momentum are radiated.

\begin{figure}[t]
   \centering
   \includegraphics[width=\columnwidth]{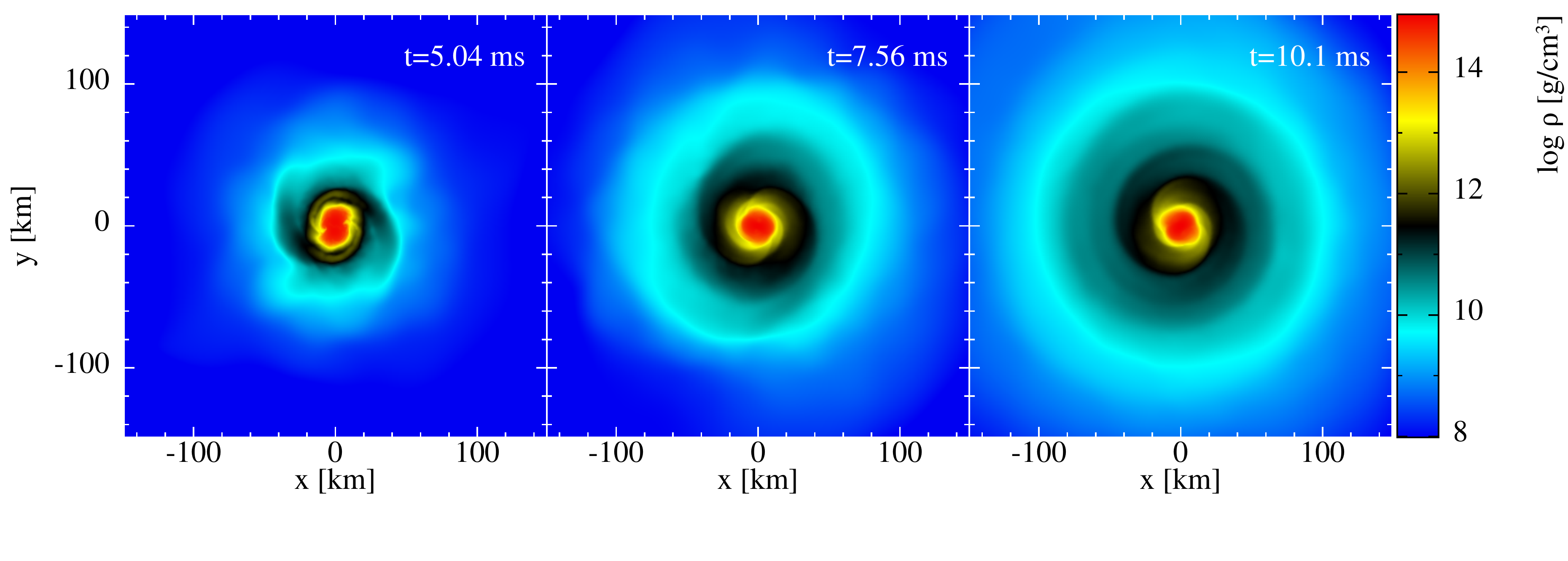} 
   \caption{Density distribution in the orbital plane of an 
   irrotational binary system with 2 $\times 1.3$ \Msun  with the APR3 EOS.}
   \label{fig:dens_APR3_2x1.3}
\end{figure}

\begin{figure}
   \centering
   \centerline{\includegraphics[width=0.5\columnwidth]{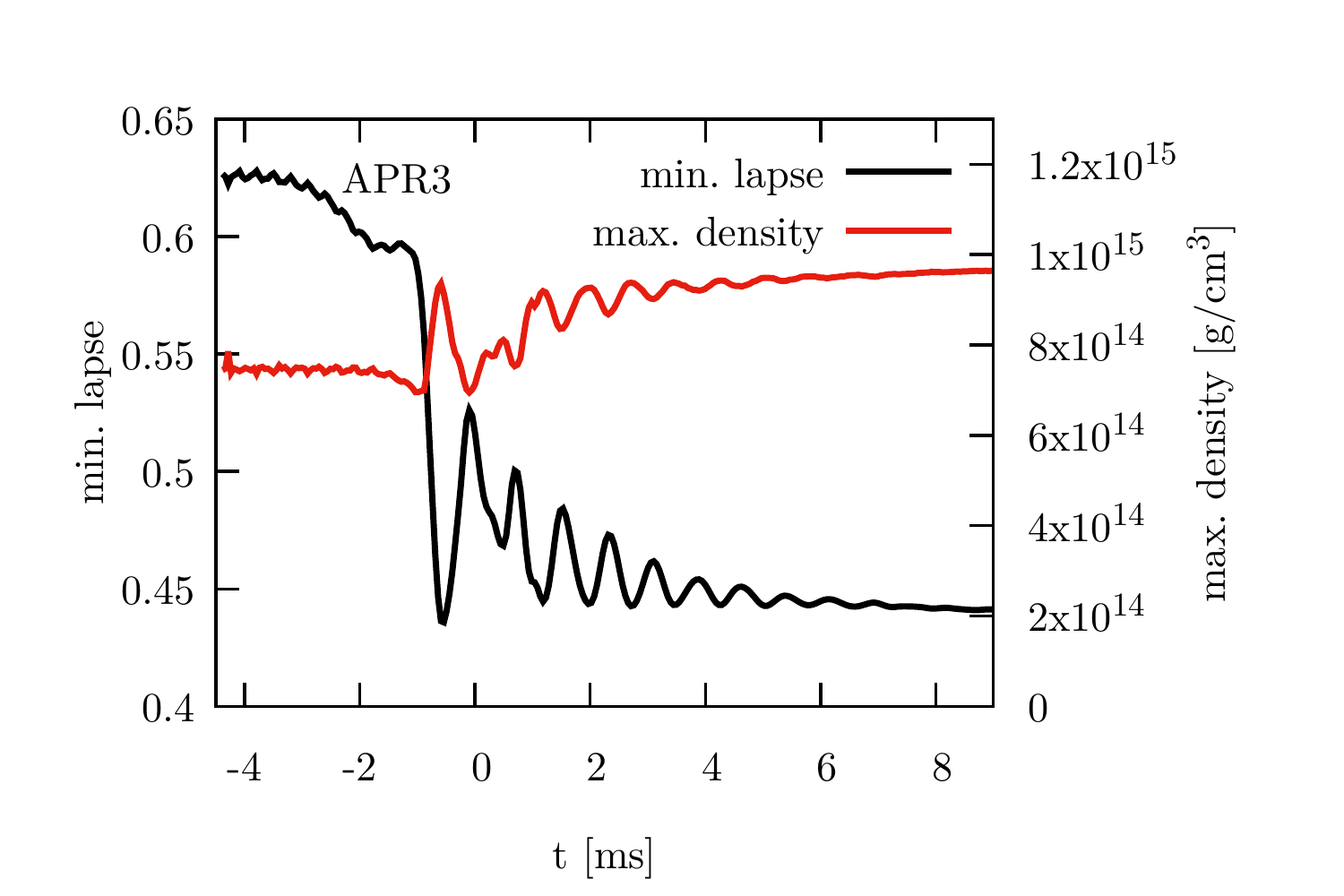} 
   \includegraphics[width=0.5\columnwidth]{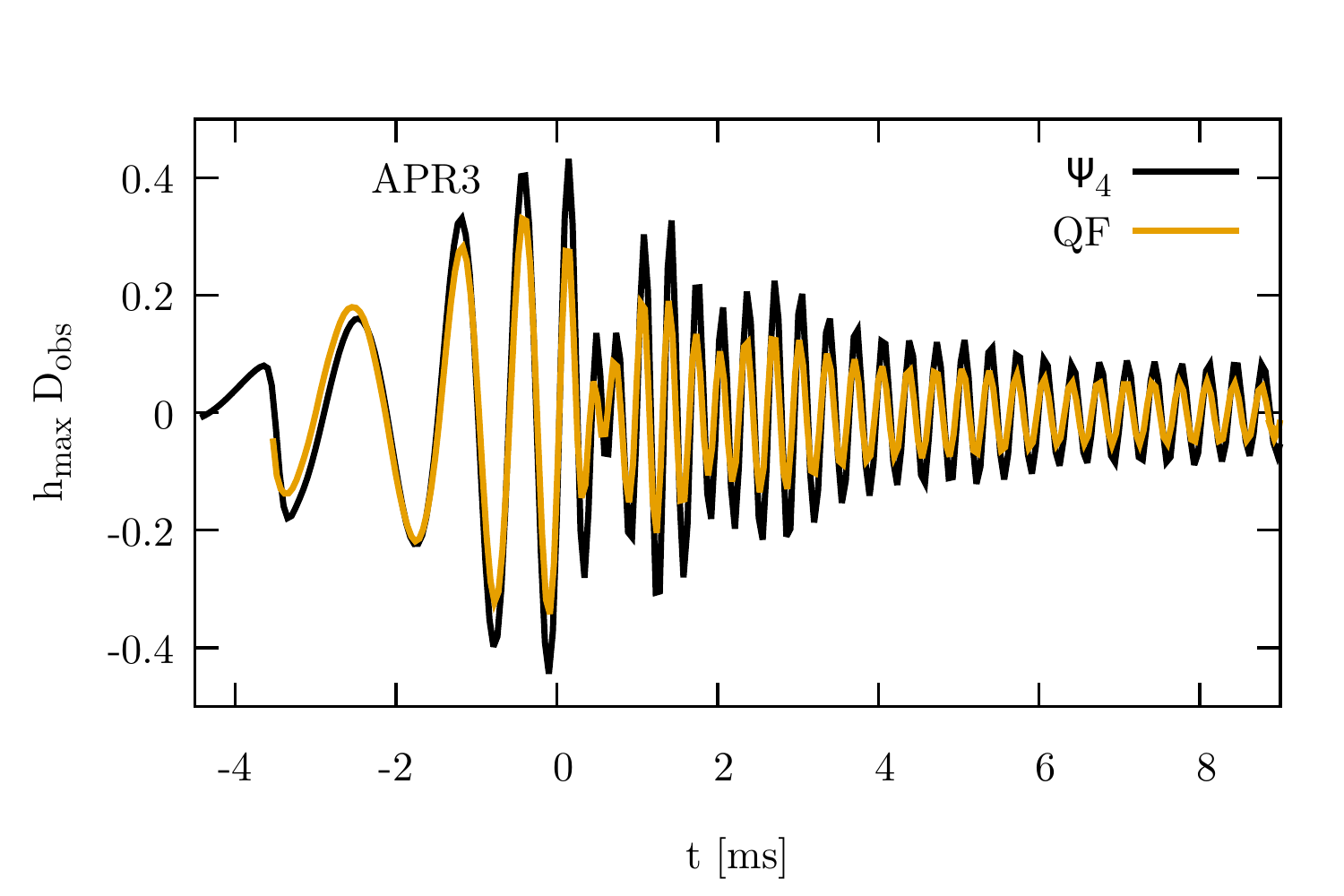}}
   \caption{{\bf (Left)} Maximum mass density (red curve, right axis) together with the minimum value of the lapse function $\alpha$ (black curve, left axis). {\bf (Right)} Maximum gravitational wave amplitude extracted via the Weyl scalar $\Psi_4$ (black curve) and the quadrupole formula (yellow curve). Both panels refer to the simulation of an irrotational 2 $\times$ 1.3 \Msun neutron star binary shown in Figure~\ref{fig:dens_APR3_2x1.3}. For convenience with comparison with other plots, $t=0$ correspond to the time of the maximum amplitude in the gravitational waveform.}
   \label{fig:lapse_APR3_2x1.3}
\end{figure}

\subsection{Neutron star merger with black hole formation}
\label{sec:BH_formation}
We also show the merger of two 1.5 \Msun neutron stars, with initial conditions produced by \lorene.
Only 1 million particles are used here (runs with higher resolution become very slow due to the timestep requirements)
with a corresponding initial grid spacing of $\Delta x=499$ m. In this case, the merged object is massive enough that 
it undergoes a prompt collapse.
During the collapse additional
grid refinement levels are added
when needed and at the end we have a total of 11 refinement levels with a finest grid resolution of
$\Delta x=32$ m.\\
In Figure~\ref{fig:dens_APR3_2x1.5} we show 3 snapshots
of the equatorial density. The first, at $t=3.28$ ms,
is from well before particles start to be removed and the density is still increasing. At $t=4.29$ ms, about 90\% of the particles have already been removed, but the maximum density of the remaining particles is still close to the initial
central density of the stars. At $5.29$ ms, matter has been drained down to $4\times 10^{-3}$ \msun. Only about $6\times 10^{-4}$ \Msun of this material is unbound from the black hole.\\
In the left panel of Figure\ref{fig:lapse_APR3_2x1.5}
we show the evolution of the maximal density (red curve, right axis) and the minimum lapse (black curve, left axis). In this plot, particles that 
have been converted to dust, do not count towards the maximum density and minimum lapse. The rapid drop in
the maximum density is completely due to the conversion of particles to dust and their eventual
removal at lapse values below 0.02.\\
In the right panel of Figure\ref{fig:lapse_APR3_2x1.5} we show a 
comparison between the maximum GW amplitude times
the distance to the observer as extracted from the quadrupole 
formula and from $\Psi_4$. As expected, the quadrupole waveform shuts off too
early as it is sourced by the matter motion and does not know about the quasinormal ringdown
of the spacetime itself. In this simulation about
$3.8\times 10^{-2}$ \Msun of energy and $1.18$ \msun$^2$ of angular momentum is radiated away by GWs.
By analyzing the properties of the final horizon using the tools from the \etk\ (\cite{Loffler:2011ay}), 
we find that the black hole that formed has a 
dimensionless spin parameter of about $a/M \approx 0.8$, consistent with earlier findings that binary neutron star 
mergers leads to faster spinning black holes than binary black hole mergers (see e.g.\ \cite{baiotti08}).
These last numbers should be taken by a grain of salt, as the finite resolution may still have an impact. A more detailed analysis is left for future work.

\begin{figure}[t]
   \centering
   \includegraphics[width=\columnwidth]{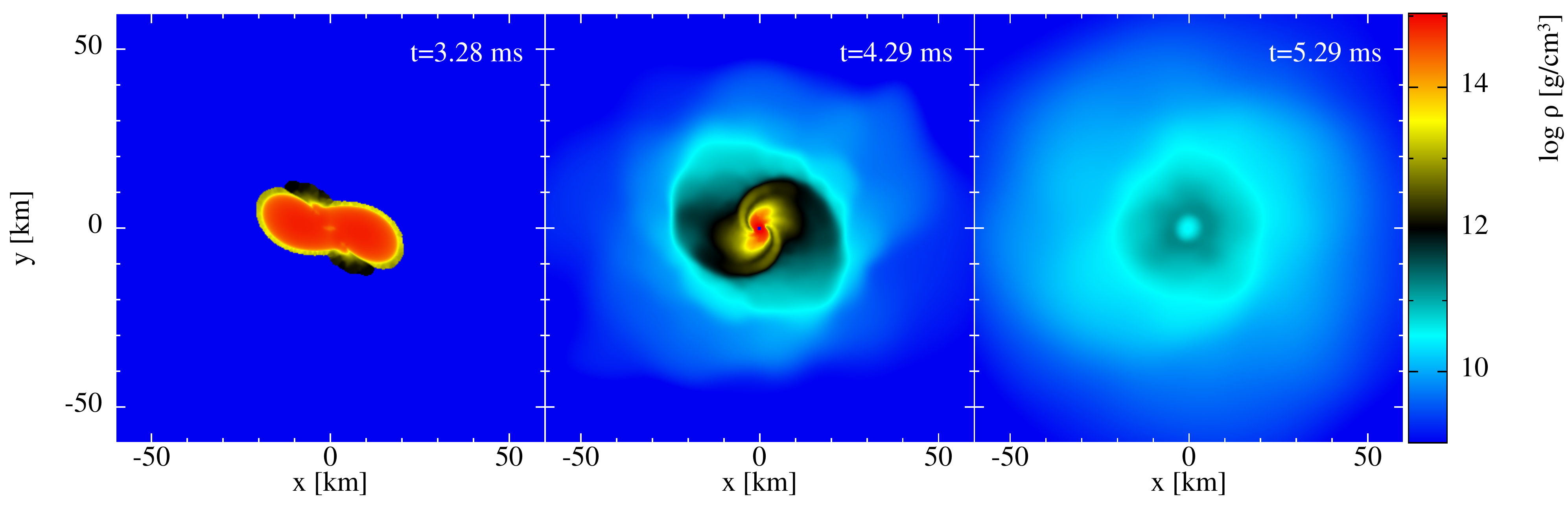} 
   \caption{Density distribution in the orbital plane of an 
   irrotational binary system with 2 $\times 1.5$ \Msun  with the APR3 EOS.}
   \label{fig:dens_APR3_2x1.5}
\end{figure}
\begin{figure}
   \centering
   \centerline{\includegraphics[width=0.5\columnwidth]{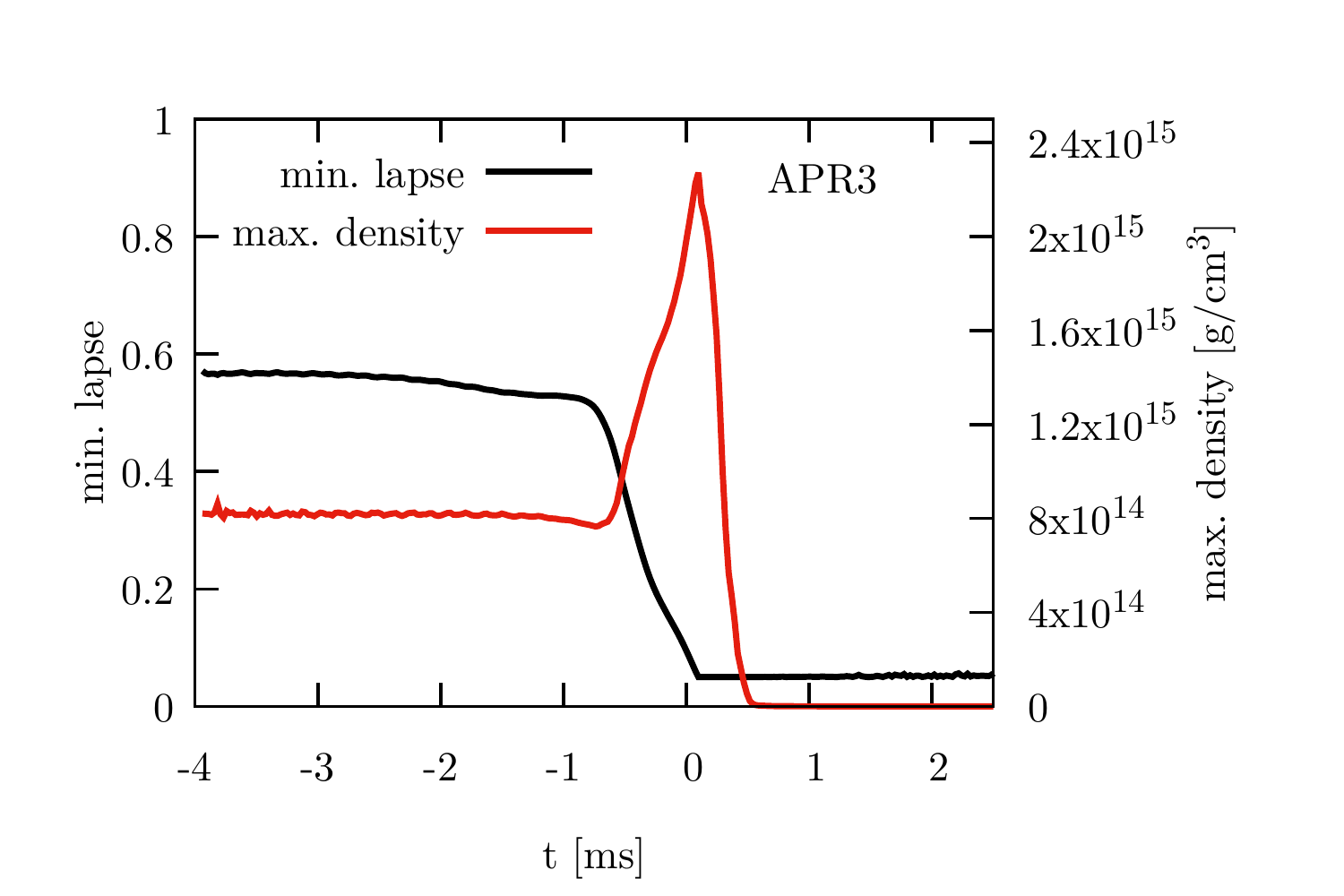} 
   \includegraphics[width=0.5\columnwidth]{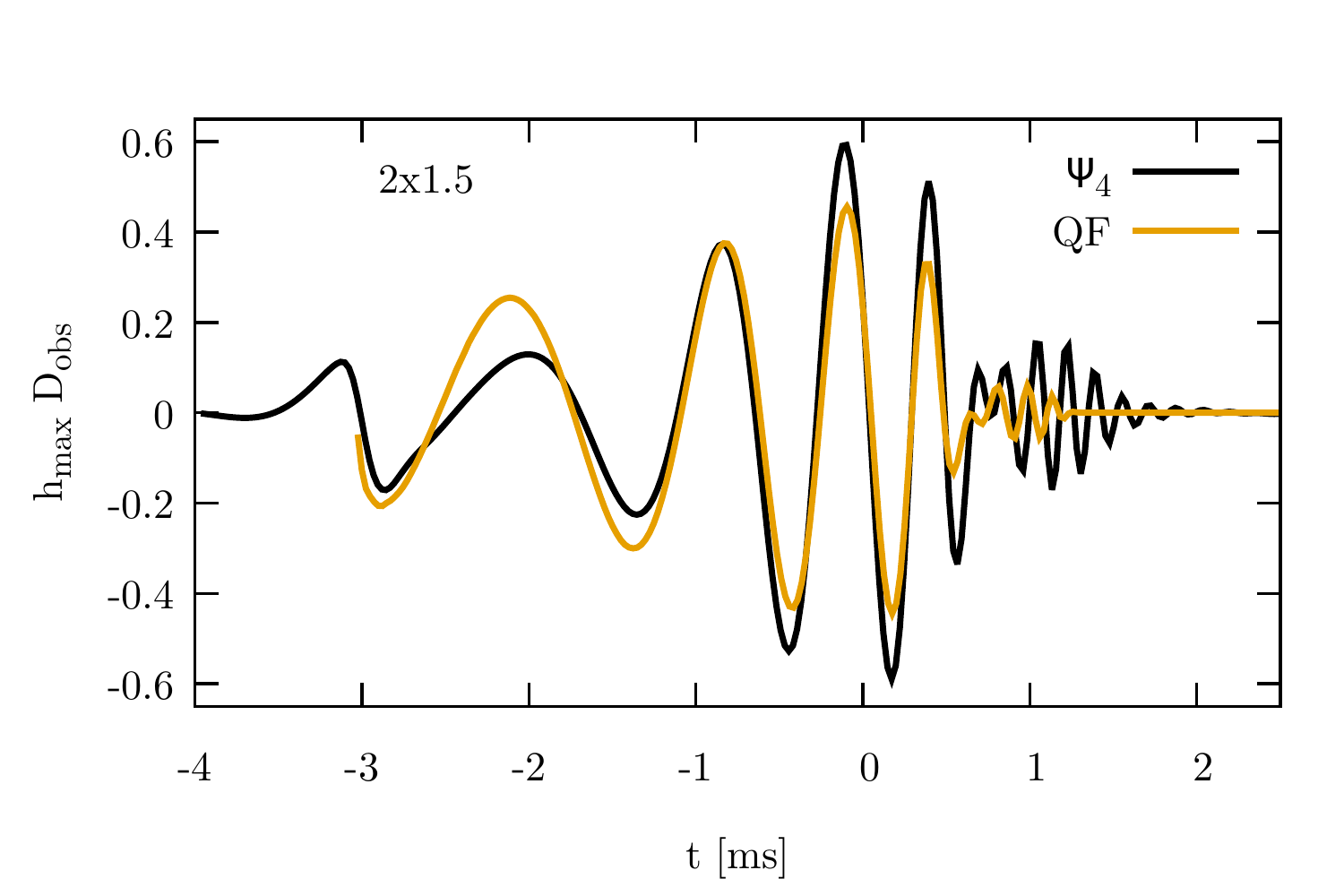}}
   \caption{{\bf (Left)} Maximum mass density (red curve, right axis) together with the minimum value of the lapse function $\alpha$ (black curve, left axis). {\bf (Right)} Maximum gravitational wave amplitude extracted via the Weyl scalar $\Psi_4$ (black curve) and the quadrupole formula (yellow curve). Both panels refer to the simulation of an irrotational 2 $\times$ 1.5 \Msun neutron star binary. For convenience with comparison with other plots, $t=0$ correspond to the time of the maximum amplitude in the gravitational waveform.}
   \label{fig:lapse_APR3_2x1.5}
\end{figure}

\subsection{Neutron star merger with a single spinning star}
\label{sec:single_Spin}
{Most commonly, irrotational binary systems are studied, and they are considered as most realistic since dissipative effects cannot spin up neutron stars to substantial spin values \citep{bildsten92,kochanek92} and, at merger, any residual stellar spin is likely small compared to the huge orbital angular momentum. Nature, however, likely can produce neutron star binary systems in several ways \citep{tauris23} and, likely at smaller rates, more extreme systems may be produced. As one such example,
we study here a binary system where only one of the two neutron stars is rapidly spinning while the other is irrotational. Such systems have hardly been explored before, we are only aware of one such study by \cite{papenfort22} where authors study extreme mass ratio and spin spin configurations.\\
We evolve a binary system with $2 \times 1.3$ \Msun stars,
where one of the stars is spinning. The chosen
value of the spin parameter, $\chi= 0.5$, corresponds to 
a spin period of about 1.2 ms.
Since \lorene cannot construct such a case, we use the \fuka library
instead. As can be seen from Figure~\ref{fig:APR3_2x1.3_chi0.5}, this initial data produces a matter distribution that is substantially different from the case shown in Section~\ref{sec:stable_remnant}. {During merger, a massive tidal tail forms and our evolution here is, qualitatively, similar to panel 1 in Figure 1 of \cite{papenfort22}. (Note however that their system has a different spin value, a different EOS and a different mass.) In panels two and three of Figure~\ref{fig:APR3_2x1.3_chi0.5} one sees how the rapidly spinning central remnant is punching shock waves into the remnant.
This strong shock compression in the torus drives polar outflows at 
$\sim 40^\circ$ from the polar axis, see the volume rendering in the left panel of  Figure~\ref{fig:PolarOutflow}, with velocities up to $\sim 0.4$ c (right panel same figure).
The density compression at merger is much milder, see left panel of Figure~\ref{fig:lapse_APR3_2x1.3_spin},
and the post-merger gravitational wave amplitudes (right panel) are substantially 
lower than in the non-spinning case. This effect is also reflected in the Fourier power spectra shown in
Figure~\ref{fig:spectra}. Here the non-spinning case (with higher power) is shown in blue and the spinning case in orange. It can
also be seen that the frequency of the main peak after merger shifts to lower frequency in the spinning case compared to the non-spinning case, consistent with the less compact remnant.\\
{As a quick test, we can compare the peak frequency with the prediction of the empirical quasi-universal relation given by Equation~(4) in \cite{Vretinaris_2020}. This relation is
\begin{equation}
f_{\mathrm{peak}}/M_{\mathrm{chirp}}=13.822-0.576 M_{\mathrm{chirp}}-1.375 R_{1.6}+0.479 M_{\mathrm{chirp}}^2-0.073 R_{1.6} M_{\mathrm{chirp}}+0.044 R_{1.6}^2,\label{eq:uni_rel}
\end{equation}
where $R_{1.6}$ is the circumferential radius of
a 1.6\Msun star with the given EOS and
$M_{\mathrm{chirp}}=(m_1 m_2)^{3/5}/(m_1+m_2)^{1/5}$ is the chirp mass
of the binary.
Using \lorene we can calculate $R_{1.6}$ for a star with the APR3 equation of state to be $R_{1.6}=11.75$ km. With $m_1=m_2=1.3$ \Msun we find
a chirp mass of $M_{\mathrm{chirp}}=1.13$ \msun.
Inserting these numbers into Equation~\eqref{eq:uni_rel}
we find that the relation predicts a value of
$f_{\mathrm{peak}}=3.09$ kHz, in excellent
agreement with the location of the peak of the spectrum for the
non-spinning case (blue curve).
Finally, in Figure~\ref{fig:dEdJ} we compare the radiated energy (red curve, left axis) and angular momentum (black curve, right axis) for the non-spinning (left panel) and spinning (right panel) case.
In both cases, the values are given as a percentage of the initial ADM value and the solid line only
includes the contribution form $\ell=2$, whereas the dashed line includes the contribution from all modes up to $\ell=4$. The non-spinning case radiates about twice as much energy as the spinning case and does it
predominantly in the $\ell=2$ modes. The spinning case does show a bit more contribution from the higher $\ell$ modes, consistent with a more asymmetric merger.\\
Since our 
main aim here is to demonstrate that
challenging astrophysical problems can be robustly 
addressed by \texttt{SPHINCS\_BSSN\_v1.0},
we leave a further discussion of the astrophysical
implications of such problems to future publications.

\begin{figure}
   \centering
   \includegraphics[width=\columnwidth]{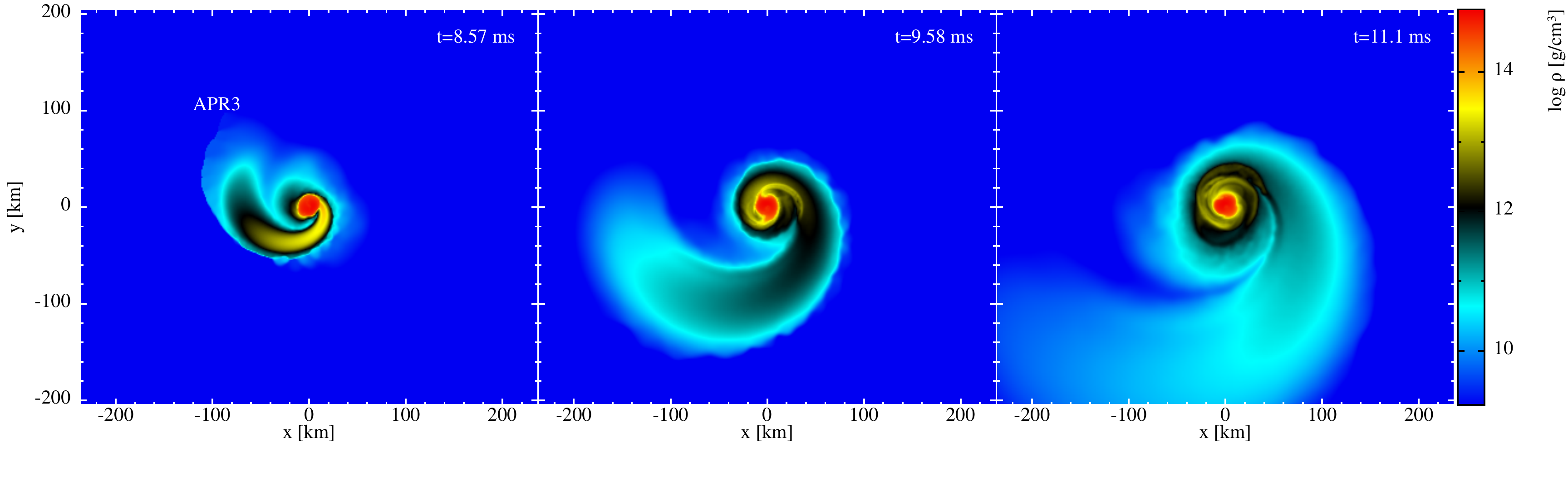} 
   \caption{Density distribution in the orbital plane of a 2 $\times 1.3$ \Msun binary with the APR3 EOS. One of the stars no spin, while the other has $\chi\simeq 0.5$.}
   \label{fig:APR3_2x1.3_chi0.5}
\end{figure}

\begin{figure}
   \centerline{
   \includegraphics[width=0.5\columnwidth]{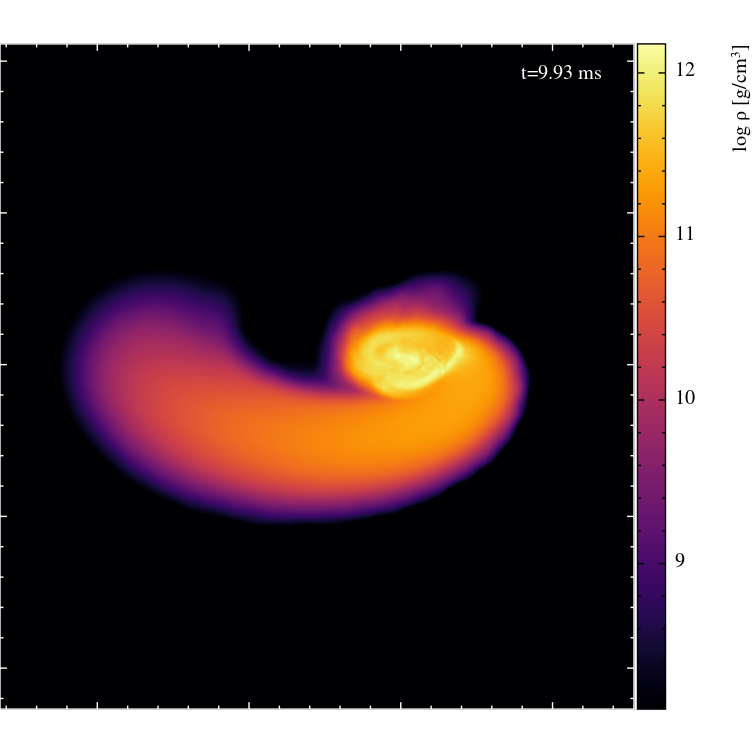} 
   \includegraphics[width=0.47\columnwidth]{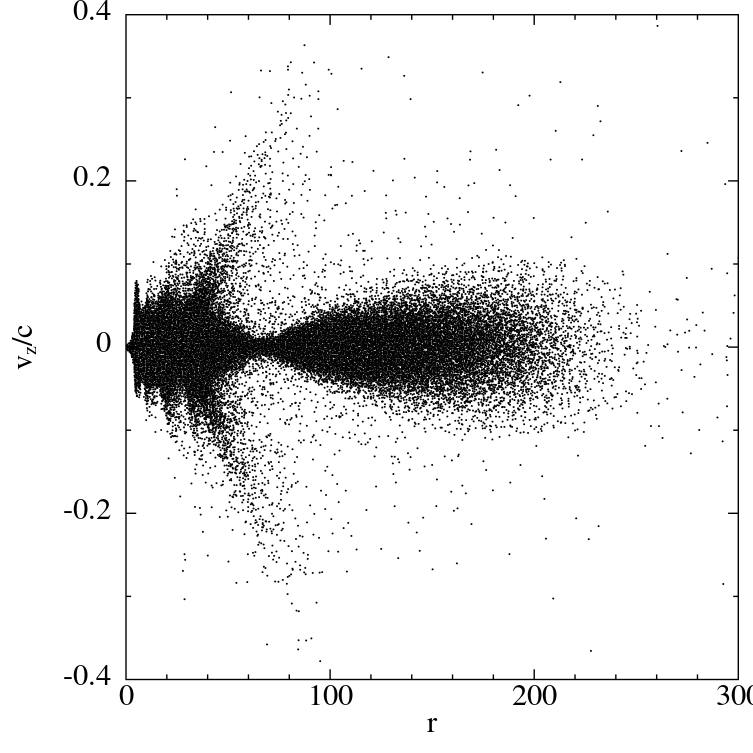} 
   }
   \caption{{\bf (Left)} Volume rendering of the density distribution at $t= 9.93$ ms of the $2 \times 1.3$ \Msun merger with a single spinning star. The rapidly spinning central object compresses the torus by
   shock waves which results in polar bulk outflows  with velocities reaching $\sim 0.4c$ {\bf (Right)}.}
   \label{fig:PolarOutflow}
\end{figure}

\begin{figure}
   \centering
   \centerline{\includegraphics[width=0.5\columnwidth]{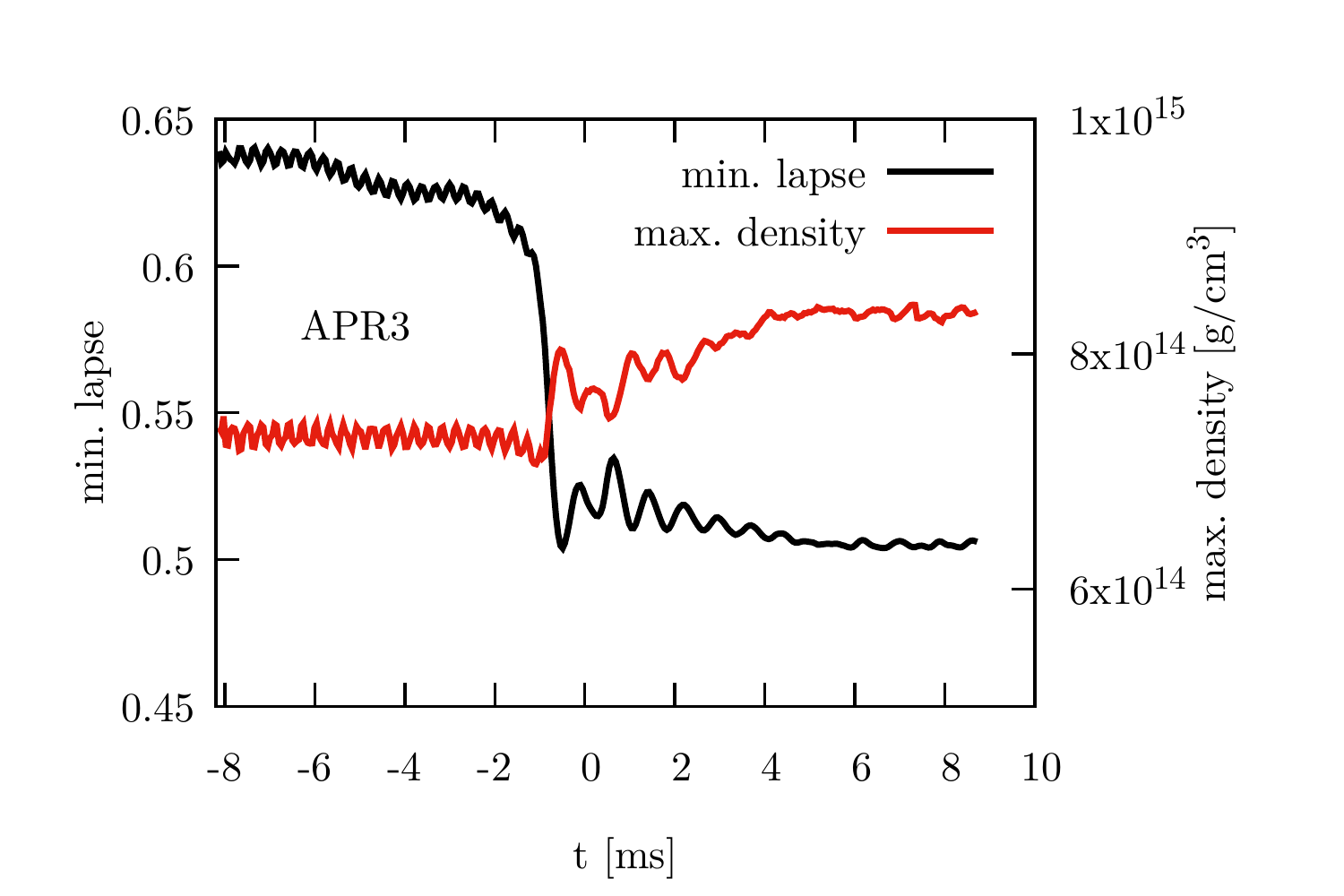} 
                \includegraphics[width=0.5\columnwidth]
                 {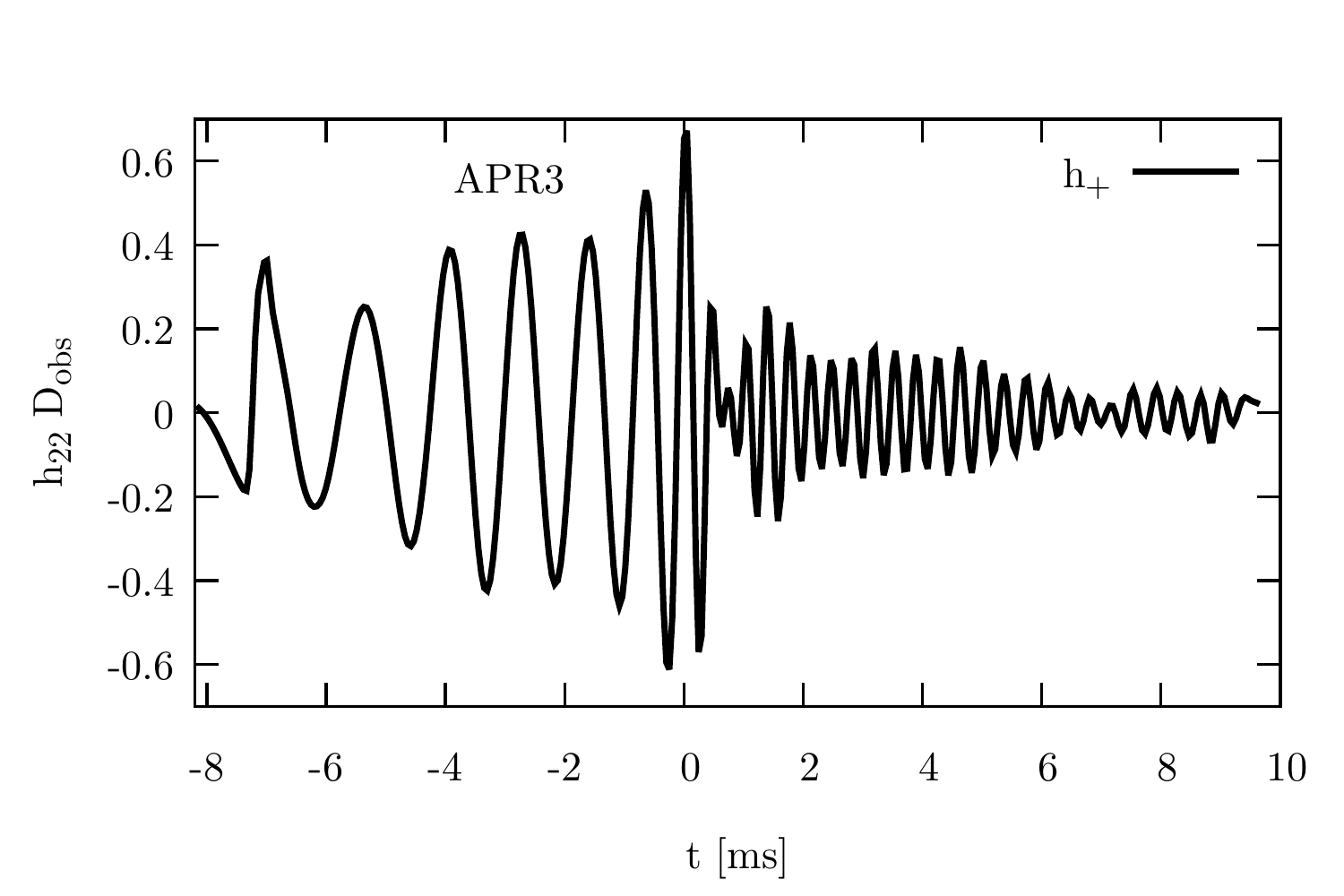}}
   \caption{{\bf (Left)} Maximum mass density (red curce, right axis) together with the minimum value of the lapse function $\alpha$ (black curve, left axis. {\bf (Right)} $\ell=2,m=2$
   mode of the $h_+$ polarization of the gravitational wave strain extracted from the Weyl scalar $\Psi_4$. Both panels refer to the simulation of a 2 $\times$ 1.3 \Msun neutron star binary where one star has a spin of $\chi\simeq 0.5$. For convenience with comparison with other plots $t=0$ correspond to the time of the maximum amplitude in the gravitational waveform.
   }
   \label{fig:lapse_APR3_2x1.3_spin}
\end{figure}

\begin{figure}
   \centering
   \centerline{\includegraphics[width=1.0\columnwidth]{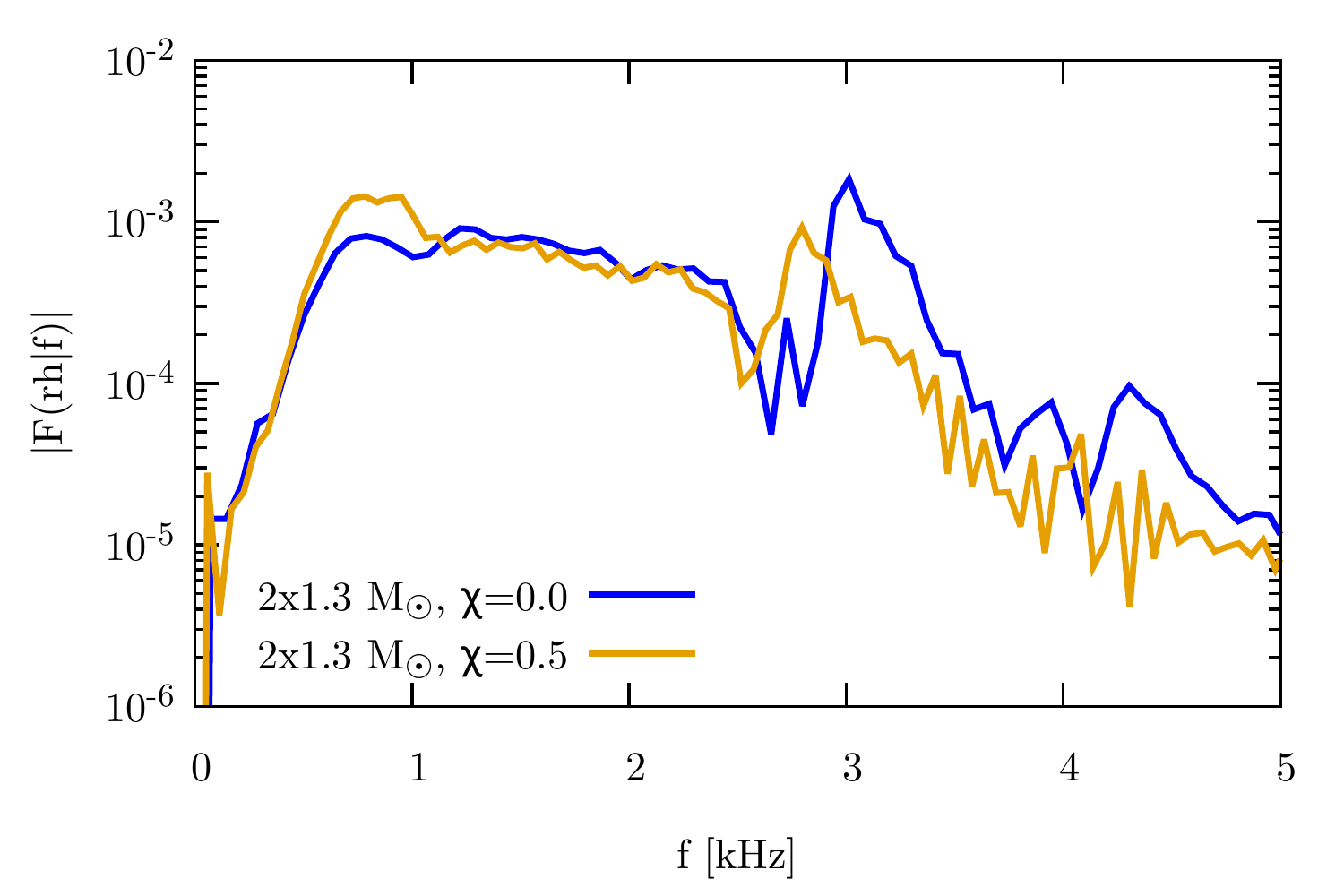}}
   \caption{Spectra of the $\ell=2$, $m=2$ gravitational waveforms for the non-spinning
   case (blue curve) and the case with one spinning star (orange curve).}
   \label{fig:spectra}
\end{figure}

\begin{figure}
   \centering
   \centerline{\includegraphics[width=\columnwidth]{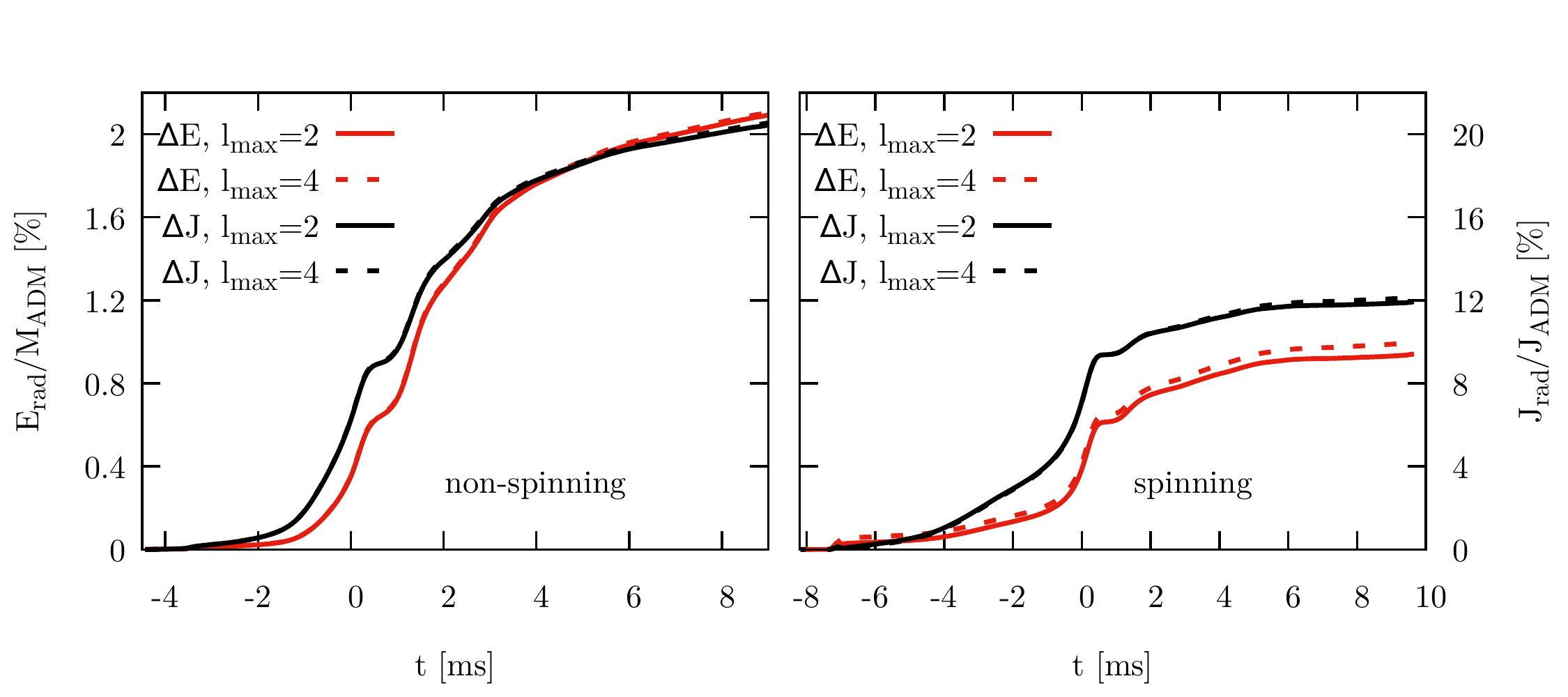}}
   \caption{The radiated energy (red curves) and angular momentum (black curves) as function of time with $t=0$ corresponding to the peak amplitude of the gravitational waveform. Both are shown as percantages of the initial ADM values. The left axis is for energy and the right axis for angular momentum. The solid lines only includes the contribution from the $\ell=2$ modes, while the dashed lines includes all modes up to $\ell=4$. The plot on the left is for the non-spinning case while the plot on the right is for the case with one spinning star.}
   \label{fig:dEdJ}
\end{figure}
\section{Summary}
\label{sec:summary}
In this work, we have presented version 1.0 of our Lagrangian Numerical Relativity code \texttt{SPHINCS\_BSSN}. 
Some of the methodological elements  have been published before \citep{rosswog21a,diener22a,rosswog22b}, others
are introduced here for the first time.\\
First, a new way to map the stress--energy tensor $T_{\mu\nu}$ (known  at the particle positions)  to our spacetime mesh is introduced. The new method sets
up a polynomial bases of a given order at each grid point and then computes
expansion coefficients that are optimal (for the given order) in the sense that
they minimize an error functional. We do this for polynomial orders from 0 to
3 and out of those possibilities we select the one that best represents the
surrounding particle values {\em and} meets some admissibility criteria. Our
procedure is described in detail in Section~\ref{sec:M_ST_coupling} and we show
an instructive example of the method in Appendix~\ref{sec:LRE_appendix}.\\
Second, we have introduced measures that make the simulation of a collapse to a black hole more robust. We realized that in some cases our originally
chosen spacetime evolution was not resolved well enough and we now add
additional refinement levels when the hydrodynamically allowed time step
drops substantially below the time step that is admissible for the space time
evolution. Once the lapse value at a particle position
has dropped to a very small value (here $\alpha_{\rm cut}= 0.02$), we remove the particle to avoid
the time step shrinking towards zero. The lapse value $\alpha_{\rm cut}$ 
is well below the value where an apparent horizon forms ($\sim 0.3$). While evolving
towards this very low lapse value, the recovery of the physical variables from
the numerical ones can fail. While this happens well inside the horizon and thus
should not affect the spacetime outside of it, we nevertheless need to keep the
particle evolution going until the threshold lapse for removal is reached. to avoid this problem, we transform the corresponding fluid particle into ``dust'' with vanishing
pressure and internal energy when the lapse at a particle drops below $a_{\mathrm{dust}}$. This allows for a simple and robust recovery of the physical variables,
and the particle's contribution to the stress--energy tensor is counted until
it is finally removed. For more details on the procedure, see Section~\ref{sec:stable_BH_formation}.\\
The third improvement concerns the placement of the SPH particles modeling the fluid at the level of the initial data, and is implemented in the code \spi. This code can now use initial data produced with the \fuka library, in addition to those produced with the \lorene library. In the latest version of the code,
the particles---both physical particles modeling the stars, and boundary particles used in the ``Artificial Pressure Method"---are placed so that they model the geometry of the stars more accurately than before. This allows for a better approximation of hydrodynamical equilibrium  with SPH particles. After their initial placement, the particles are
iterated into optimal positions according to a variant of the Artificial Pressure Method.
In the original version of this method, the relative error between the density provided by the ID solver and the SPH estimate, was used to define an
``artificial pressure." The latter's gradient pushes the particles in positions where they reduce
the error on the density. In the latest version of this method, we instead use the relative error of the {\em physical} pressure (rather than the density) to compute the artificial pressure. 
This minimizes the error on the physical pressure directly, and leads (with everything else being the same) to lower errors in the physical pressure
and thus to more accurate initial data.\\
To illustrate the working and robustness of \spB1 we have performed three simulations:
one irrotational binary merger ($2 \times 1.3$ \msun) that remains stable on the simulation timescale, one irrotational system  ($2 \times 1.5$ \msun)
that collapses ``promptly" (i.e. without any bounce) and one extreme binary system
where only one of the stars has a (large) spin, $\chi = 0.5$. All these simulations
use the APR3 equation of state, the first two simulations are produced using \lorene,
the latter using \fuka.\\
Not too surprisingly, for the stable irrotational case we find an anti-correlation between the maximum density
and the minimum lapse value, see Figure~\ref{fig:lapse_APR3_2x1.3}, left panel. Concerning the GW emission, we have rather good agreement between
the quadrupole waveform (for more details see \cite{rosswog22b}) and the
waveform extracted from the Weyl scalar
$\psi_4$, see Figure~\ref{fig:lapse_APR3_2x1.3}, right panel. We show the agreement for the first two cases only, but it is similarly good for the third case. Again expected,
we find that the GW emission is strongly dominated by the $l=2, m=2$
mode. We find that GWs carry away about 2\% of the
initial ADM mass and about 20\% of the initial ADM
angular momentum.\\
For the collapsing system, we find that very little mass $<6\times 10^{-4}$\Msun escapes the fate of falling into the BH
and that the final BH is spinning fairly fast.
The dimensionless spin parameter of $a/M\approx 0.8$ is
significantly larger than the end result of an irrotational binary BH merger where $a/M\approx 0.68$.\\
Last, but not least, we performed a simulation of an extreme case with only one rapidly spinning star that has been produced using the \fuka library. We find that the neutron
star spin has a very large impact on the merger morphology. Similar to cases with extreme mass ratios, a single puffed up tidal tail forms. Overall, the collision is less
violent in the sense that the high-density regions become not as much compressed as
in the equal mass case and the minimum lapse values remain larger. We also observe
that less energy and angular momentum are radiated by GWs in the post-merger phase, likely because the central regions are less perturbed in the
less violent collision and thus deviate less from rotational symmetry.\\
Clearly,  \spB1 would benefit from the inclusion of more microphysics and 
its computational performance needs to be further improved. These issues will be addressed in future work.

\section*{Data availablity statement}
The data underlying this article will be shared on reasonable request to the corresponding author.

\section*{Author Contributions}
SR has developed the methods in and coded the fluid part of \texttt{SPINCS\_BSSN v1.0}. He has further
designed various versions of mapping particle properties to the mesh, the latest of which is described here in Section~\ref{sec:M_ST_coupling} and exemplified in Appendix 4. All of this has happened in close coordination with PD. SR has further written the first draft of this paper.\\
FT developed \spi, produced the initial data used in the simulations using \lorene, \fuka and \spi, and performed the computation of the ADM momentum of the fluid in SPH. FT wrote the part of Section~\ref{sec:hydro} describing the SPH estimate of the ADM momentum, Section~\ref{sec:id}, Appendix~\ref{app:ovals}, Appendix~\ref{app:admmomsph}, Appendix~\ref{app:admmomlorene}. All authors contributed to manuscript revision, read, and approved the submitted version.\\
PD has developed the methods necessary for grid structures (including refinement) and coded the interface to the needed routines from \McL from the
\etk and implemented them in \texttt{SPINCS\_BSSN v1.0}. PD derived and implemented the Hermite polynomial routines used for mapping metric information
from the grid to the particles and worked closely with SR on developing and testing the methods to map
the stress energy tensor from the particles to the grid.

\section*{Funding}
SR has been supported by the Swedish Research Council (VR) under 
grant number 2020-05044, by the research environment grant
``Gravitational Radiation and Electromagnetic Astrophysical
Transients'' (GREAT) funded by the Swedish Research Council (VR) 
under Dnr 2016-06012, by which also FT has been supported, and by the Knut and Alice Wallenberg Foundation
under grant Dnr. KAW 2019.0112,
by the Deutsche Forschungsgemeinschaft (DFG, German Research 
Foundation) under Germany’s Excellence Strategy – EXC 2121 
"Quantum Universe" – 390833306 and by the European Research 
Council (ERC) Advanced Grant INSPIRATION under the European 
Union’s Horizon 2020 research and innovation programme 
(Grant agreement No. 101053985). 
SR's calculations were performed 
on the facilities of the North-German Supercomputing Alliance (HLRN), 
on the resources provided by the Swedish National Infrastructure for 
Computing (SNIC) in Link\"oping, partially funded by the Swedish Research 
Council through Grant Agreement no. 2016-07213, and at the SUNRISE HPC 
facility supported by the Technical Division at the Department of 
Physics, Stockholm University.


\section*{Acknowledgments}
We thank Ian Hawke for insightful comments on our LRE method and we are very grateful to Sam Tootle for his help with the \fuka library.
We also thank the referees for insightful and useful comments.
 Special thanks go to Holger Motzkau 
and Mikica Kocic for their excellent support in upgrading and 
maintaining SUNRISE.

\section*{Conflict of interest}
The authors declare that the research was conducted in the absence of any commercial or financial relationships that could be construed as a potential conflict of interest.

\section*{Publisher's note}
All claims expressed in this article are solely those
of the authors and do not necessarily represent those of
their affiliated organizations, or those of the publisher,
the editors and the reviewers. Any product that may be
evaluated in this article, or claim that may be made by
its manufacturer, is not guaranteed or endorsed by the
publisher.



\bibliographystyle{Frontiers-Harvard} 
\bibliography{biblio,astro_SKR_fixed}

\appendix
\renewcommand*{\thesection}{\Alph{section}}
\renewcommand*{\thefigure}{A\arabic{figure}}
\setcounter{figure}{0}
\section*{Appendices}

\input{sphincs_id-appendices}

\section{An example of function approximation via a Local Regression Estimate (LRE)}
\label{sec:LRE_appendix}
To illustrate the function approximation via LRE, we place a set of ``grid points" on a Cartesian mesh in $[-0.5:0.5]^3$ with a spacing of $\Delta_{\rm g}= 0.07$. Our ``particles" 
are originally placed on a Cartesian grid between $[-0.9:0.9] \times [-0.9:0.9] \times [-0.9:0.9]$ with a spacing of 
$\Delta_{\rm p}= 0.1$, but then slightly randomized by adding to each particle position component 
a random number in $[-0.01,0.01]$. This regular but not perfect distribution is meant to mimmick an SPH
particle distribution. Each of the particles is assigned a  smoothing length of $h=0.1$ and
a function value according to 
\be
f(x,y,z)= 10 - \sqrt{x^2 + y^2 + z^2}.
\ee
We then calculate the LRE approximations of a specified polynomial order at each grid point
and we measure at the grid points the relative error of the LRE estimate compared with the original
function $f$, see Figure~\ref{fig:func_err}, left panel.
\begin{figure*} 
   \centering
   \centerline{\includegraphics[width=\textwidth]{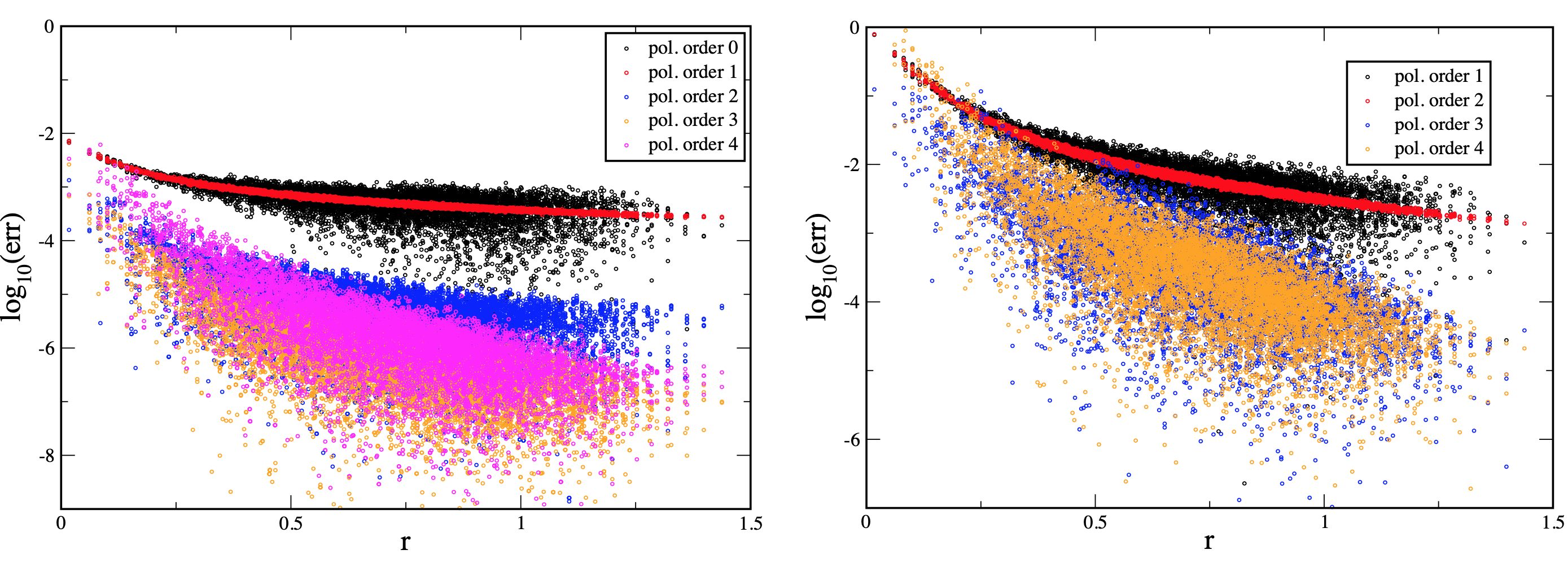}}
   \caption{Relative errors of the function approximation {\bf (left)} and the gradient of the function approximation {\bf (right)}
    as function of radius of the grid points.}
   \label{fig:func_err}
\end{figure*}
The lowest polynomial order already provides estimates below $1\%$ error (black), but with a large spread of accuracies,
while the linear order provides estimates of the same order, but with substantially less noise (red). The transition from linear
to quadratic polynomials (blue) provides a serious enhancement of the accuracy, while cubic (orange) and quartic polynomials 
(magenta) only marginally improve the results further.\\
The situation is similar for the gradient estimates, where we use
\be 
{\rm err}\equiv \frac{| \; |(\nabla f)_{\rm num}| - |(\nabla f)_{\rm theo}| \; |}{|(\nabla f)_{\rm theo}|}.
\ee
as error measure and $(\nabla f)_{\rm num}$ is the numerically calculated gradient estimate while $(\nabla f)_{\rm theo}$
is the exact result. Also here, the overall scale between the lowest (linear) and next-to-lowest order (quadratic) is similar,
but the quadratic results are much less noisy. Again we find a substantial improvement going to the next order (cubic), but
no more substantial gain when further increasing the polynomial order.\\
\begin{figure*} 
   \centering
   \centerline{\includegraphics[width=1.05\textwidth]{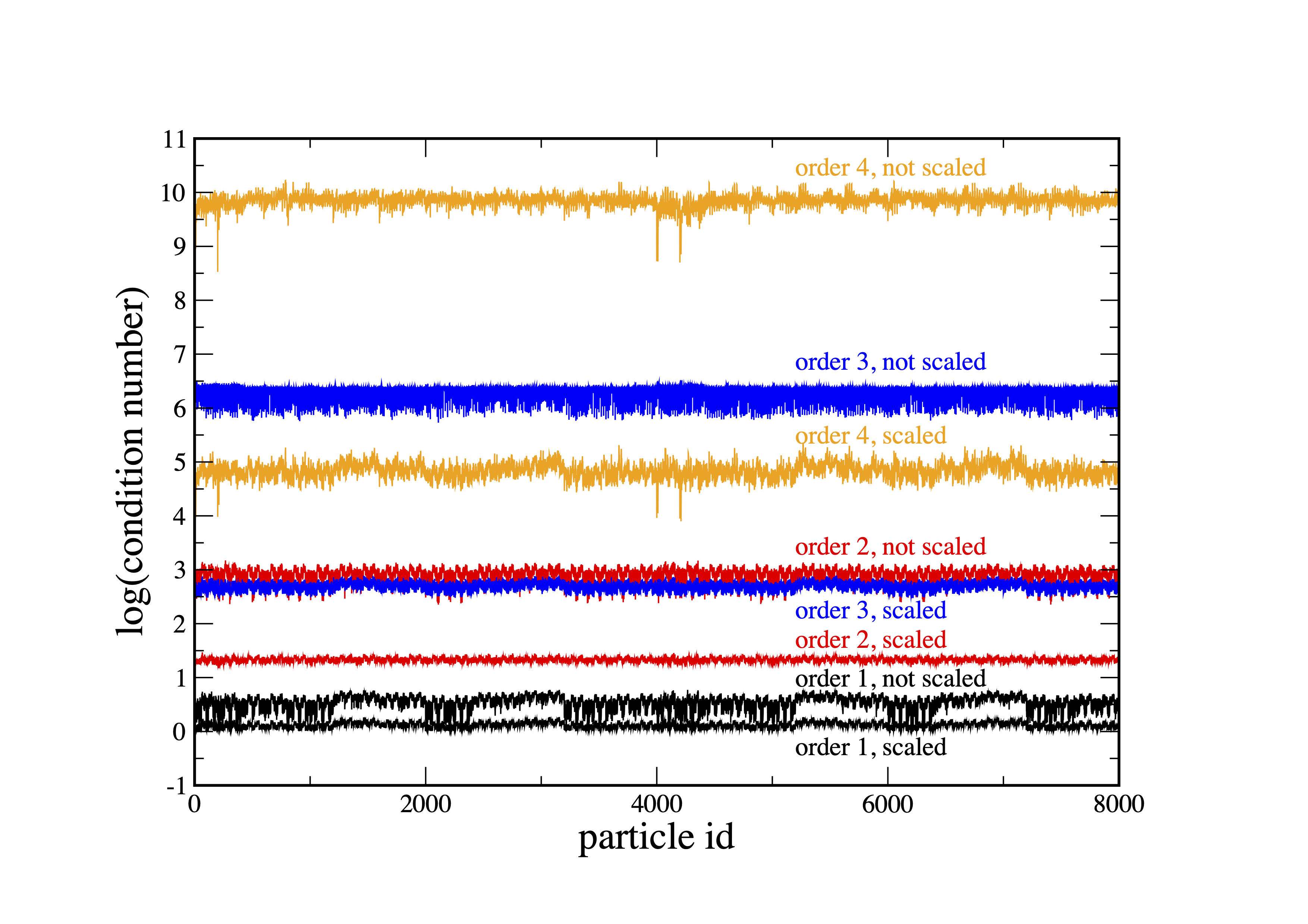}}
   \caption{Condition number for (not) re-scaled basis functions for polynomial
   orders from one to four. Note that for order four the re-scaling improves the condition
   number by as much as five orders of magnitude.}
   \label{fig:condition_number}
\end{figure*}
{We also want to briefly illustrate the effect of re-scaling the moment matrix with appropriate powers of the length $l_p= (\nu_p/N_p)^{1/3}$ to ensure
that all matrix elements have the same dimension.
In practice, this is achieved by de-dimensionalizing
the shifted basis functions, e.g. $\Delta^{\rm G}x \rightarrow \Delta^{\rm G}x/l_p$, $\Delta^{\rm G}x 
\Delta^{\rm G}y \rightarrow \Delta^{\rm G}x \Delta^{\rm G}y/l_p^2$ etc. We show in Figure~\ref{fig:condition_number} the condition numbers, $\mathcal{C}\equiv ||M M^{-1}||$, a measure for how close a matrix is to being singular, for the above experiment, once for the straight forward and once for the re-scaled version. The condition numbers in the re-scaled case improve dramatically, e.g. by five orders of magnitude in the case of quartic basis functions (orange curves in Figure~\ref{fig:condition_number}). We do, however, not see any noteworthy improvement of the accuracy which we interpret as a success of the well-working Singular Value Decomposition. Nevertheless, since the re-scaling is essentially free of computational cost,
we always use re-scaled basis functions. 

\end{document}

%% file: preamble.tex
\usepackage{verbatim}

\usepackage{amsmath,amssymb,amsopn,amstext,amsthm,mathtools}
\usepackage[scr=kp]{mathalpha}

\usepackage[normalem]{ulem}

\usepackage[section]{placeins}

\usepackage{xspace}

\newcommand{\SpB}{\texttt{SPHINCS\_BSSN}\xspace}
\newcommand{\spB}{\texttt{SPHINCS\_BSSN}}
\newcommand{\Ma}{\texttt{MAGMA2}\enskip}

\newcommand{\lorene}{\texttt{LORENE}\xspace}
\newcommand{\fuka}{\texttt{FUKA}\xspace}

\newcommand{\McL}{\texttt{McLachlan}\xspace}

\newcommand{\spi}{\texttt{SPHINCS\_ID}\xspace}

\usepackage{caption}
\usepackage[toc,page]{appendix}

\global\long\def\ee{\mathrm{e}}

\global\long\def\dd{\mathrm{d}}
\global\long\def\sphS{S}
\global\long\def\eulS{s}

%% file: sphincs_id.tex
In this section, we describe recent improvements in setting up the SPH particles in the 
initial data (ID) code \spi \cite{sphincsid}. It can now also be linked to our fork 
of \fuka \cite{Papenfort_2021,kadath}---extended to comply with our needs---to produce 
BSSN and SPH ID for neutron star binaries. The \fuka codes are built on an extended version of the KADATH library \cite{GRANDCLEMENT20103334}. In this section, we refer generically to 
\lorene \cite{Gourgoulhon_2001,Grandclement_2001,lorene} and \fuka with the term ``ID solver," when the discussion applies to both solvers.

\subsection{Modeling neutron stars with the Artificial Pressure Method}
\label{subsec:admpressure}

As described in \cite[Section~2.2.2]{diener22a}, the initial neutron stars are modelled  by placing the
SPH particles according to the ``Artificial Pressure Method" (APM) which uses the 
solutions found by
the ID solver. We briefly summarize the original method below, and refer the reader to \cite{rosswog20a,rosswog21a} and \cite[Section~2.2.2]{diener22a} for more details, before we describe an additional improvement. \\
First,  particles are placed according to a freely specified geometry (lattice, spherical surfaces, etc.), and then each particle $a$ is assigned the same baryon number $\nu_0= \nu_{\rm tot}/N_{\rm part}$, 
where $\nu_{\rm tot}$ is the total baryon number for the star and $N_{\rm part}$ the number of SPH particles used to model it.
Subsequently an "artificial pressure" is defined as
\begin{align}
\label{eq:artpressure}
\pi_a= {\rm max}\left(1 + \dfrac{N_a-N_\mathrm{ID}(\vec{r}_a)}{N_\mathrm{ID}(\vec{r}_a)}, 0.1 \right).
\end{align}
Here $N_a$ is the SPH estimate of the density variable defined in Equation~\eqref{eq:N_def} on particle $a$ and $N_\mathrm{ID}(\vec{r}_a)$ is the result from the ID solver. The lower bound of 0.1 is imposed to avoid negative values.  
The major goal of the original APM is to minimize the difference between $N_a$ and $N_\mathrm{ID}(\vec{r}_a)$ while using SPH particles of the same baryon number.\footnote{Another goal of the APM is to produce a locally isotropic particle distribution \cite{rosswog20a}, that is, a distribution such that for every particle, there exist a small enough neighborhood around it such that the particle distribution inside such neighborhood is isotropic.} Therefore, at each iteration of the APM, the particle positions are updated in order to achieve vanishing
(artificial) pressure gradients. The corresponding position update reads:
\begin{align}
{\delta \vec{r}_a}^{\rm APM}= -\frac{1}{2}h_a^2\,\nu_a \sum_{b} \frac{\pi_a+\pi_b}{N_b} \,\nabla_a W_{ab}(h_a),
\end{align}
see Section~\ref{sec:hydro} for the meaning of the involved quantities. The iteration stops when the differences between $N_a$ and $N_\mathrm{ID}(\vec{r}_a)$ do not change significantly anymore.\footnote{Currently, we exit the APM iteration if the baryon number ratio $\nu_\mathrm{max}/\nu_\mathrm{min}$ does not change more than $0.25\%$ for 300 iterations. The maximum and minimum are taken over all the particles.}

\begin{figure*}[t]
    \centering
    \includegraphics[width=\linewidth]{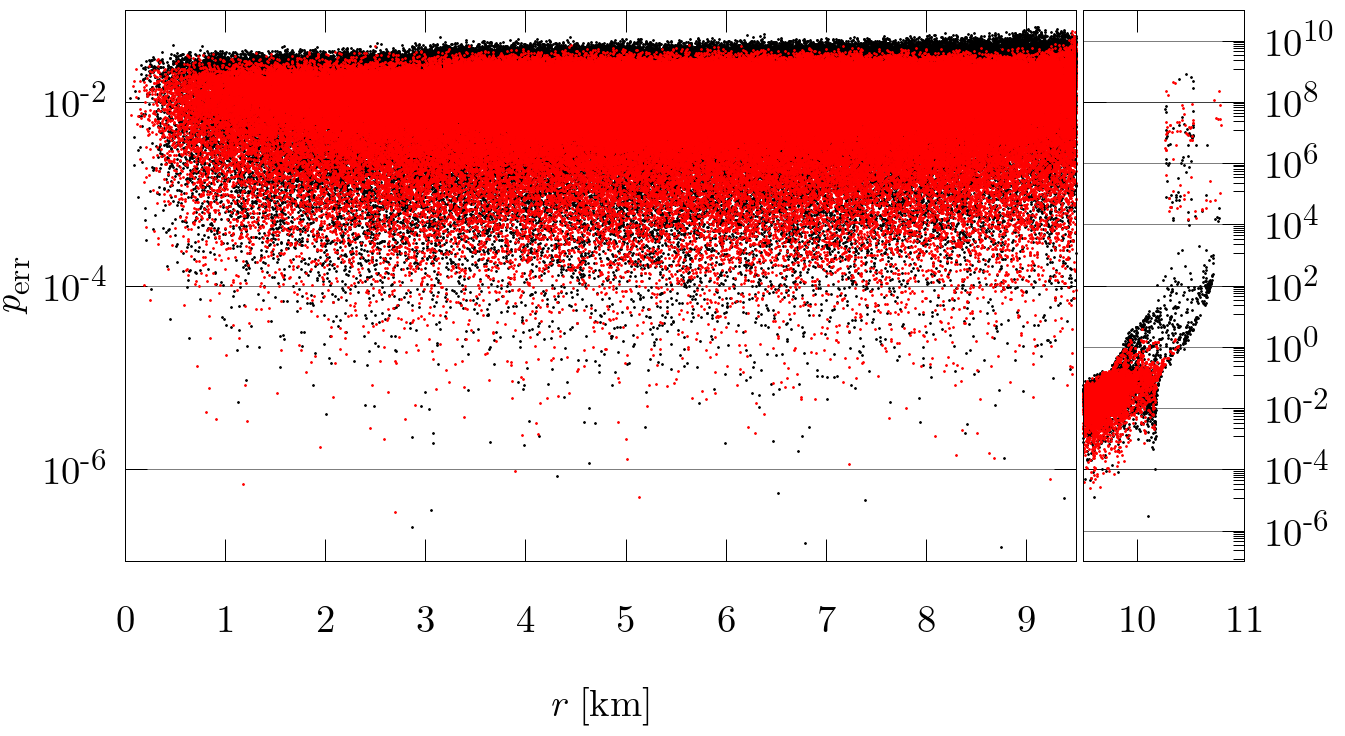} \\
    \includegraphics[width=\linewidth]{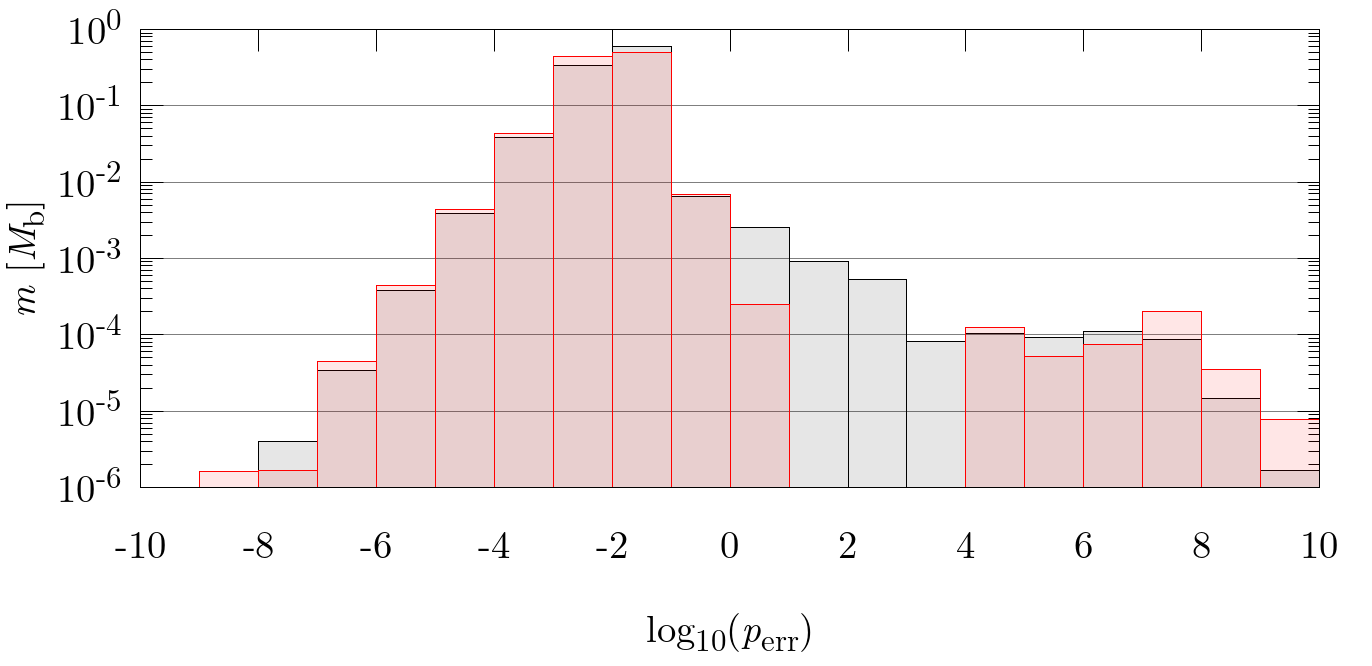}
    \caption{Relative errors on the pressure $p_\mathrm{err}\coloneqq |p(\rho_a)-p(\rho_\mathrm{ID}(\vec{r}_a))|/p(\rho_\mathrm{ID}(\vec{r}_a))$ at the end of an APM iteration that uses definition \eqref{eq:artpressure} (black) and another APM iteration that uses definition \eqref{eq:artpressure2} (red). The errors are evaluated on $2.5\times 10^6$ particles modeling the same neutron star, which belongs to a $1.9 M_\odot$ equal-mass, $47.5$ km separation, MS1b, irrotational BNS produced with \lorene. {\bf (Top)} $p_\mathrm{err}(r)$, with $r$ radial coordinate of each particle measured from the barycenter of the star. The left panel shows the data for $r\in [0,9.5]$ km; the right panel for $r\in [9.5,11]$ km. {\bf (Bottom)} $p_\mathrm{err}$ binned in log scale; each bin is one order of magnitude wide. $m$ is the sum of the baryonic masses of the particles in each bin, in units of the baryonic mass of the star $M_\mathrm{b}=2.016 M_\odot$. Except for the outermost particles with $r\gtrsim 10$ km $\simeq 93\%R$ ($R$ being the larger radius of the star), whose errors do not decrease, the errors for the particles in the bulk of the star decrease using definition \eqref{eq:artpressure2}. For example, $\sim 10\%$ of the baryonic mass of the star improves the error by roughly one order of magnitude, moving from bin $[10^{-2},10^{-1}]$ to bin $[10^{-3},10^{-2}]$.}
    \label{fig:apm-err1919}
\end{figure*}

While we  want to construct initial conditions
with densities $N_a$ as close as possible to 
$N_\mathrm{ID}(\vec{r}_a)$, it is in the end the (physical) pressure 
gradients that, apart from gravity,
drive the physical fluid motion. Therefore, it may be advantageous
to construct the {\em artificial} pressure, $\pi_a$, 
from  the {\em physical} pressures rather
than  the densities as in Equation~(\ref{eq:artpressure})
%
For a short motivation as to why to use the pressure, we will briefly
switch to a Newtonian description (the GR case with our conventions follows in a straight-forward way) and
we define, for a general EOS,  the quantity
\be
\bar{\gamma}\equiv \left(\frac{\partial P}{\partial \rho}\right)_s \frac{\rho}{P},
\ee
which, for the special case of a polytropic EOS, simply reduces to the polytropic exponent $\gamma$.
If we assume that we have found a numerical
solution for the density that is $\rho= \rho_0 + \delta \rho$,
where $\rho_0$ is the true solution, the resulting pressure
is
\be
P(\rho_0 + \delta \rho)\approx P(\rho_0) + \left(\frac{\partial P}{\partial \rho}\right)_s \delta \rho.
\ee
With $P_0=P(\rho_0)$, the relative error $\epsilon_P$ in the pressure reads
\be
\epsilon_P= \frac{P(\rho_0+\delta \rho) -P_0}{P_0}= \frac{(\partial P/\partial \rho)_s \delta \rho}{P_0}=
\bar{\gamma} \frac{\delta \rho}{\rho}= \bar{\gamma}  \; \epsilon_\rho.
\ee
Neutron star equations of state
can be approximated by piecewise polytropes \citep{read09},
which even at the lowest density piece 
(as is the case for all other equations of state that we are aware of) have a polytropic exponent value of $\gamma > 1.3$. The higher density pieces have substantially larger values, often larger than $3$. Thus, we expect that for all physically relevant EOSs (not just
polytropes) the relative error in $P$ will be larger than in $\rho$. This motivates us to change the definition of the artificial pressure 
from \eqref{eq:artpressure} to
\begin{align}
\label{eq:artpressure2}
\pi_a= {\rm max}\left(1 + \dfrac{p(\rho_a)-p(\rho_\mathrm{ID}(\vec{r}_a))}{p(\rho_\mathrm{ID}(\vec{r}_a))}, 0.1 \right),
\end{align}
so that the APM iteration minimizes the error on the pressure directly. 
In Figure~\ref{fig:apm-err1919} we see a comparison between the errors on the pressure when using the definitions \eqref{eq:artpressure} and \eqref{eq:artpressure2}. With definitions \eqref{eq:artpressure2} the errors decrease in the inner 93\%
of stellar radius, but do not improve significantly in the outermost layers.
This is the case because at finite numerical resolution, the extremely steep (physical) gradients in the stellar surface cannot be
resolved. However, these outer layers only constitute a very small fraction of the baryonic mass of the star, about $0.01\%$ for the star in Figure~\ref{fig:apm-err1919}, and are therefore not a matter of concern here.
\FloatBarrier
In summary, the errors do not reduce very significantly, 
but the change in the definition of the artificial pressure 
is nevertheless a welcome enhancement that complements
the other improvements to \spi and \spB1.
The \spi code lets the user choose which definition to use,  \eqref{eq:artpressure} or \eqref{eq:artpressure2}. The ID used for the runs described in this paper were produced using \eqref{eq:artpressure2}.

\subsection{The initial particle distribution on surface-conforming ovals}
\label{subsec:ovalsurfaces}
 
\begin{figure*}[t]
    \centering
    \includegraphics[width=\textwidth]{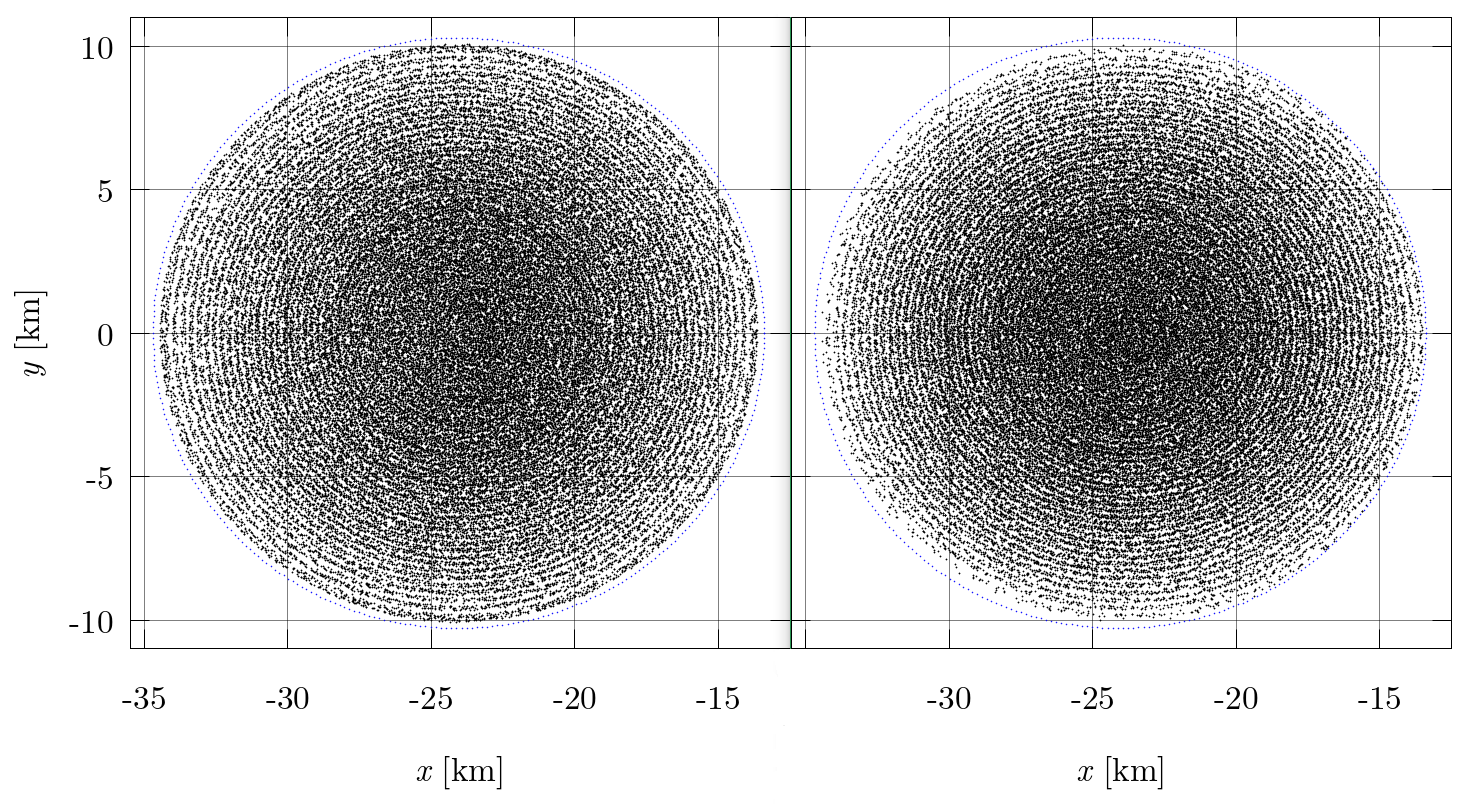}
    \caption{Cuts along the $xy$ plane, with $|z|<\,\sim\!\! 0.89$ km, of particle distributions with $10^6$ particles modeling the same star. The left panel shows particles placed on surface-conforming ovals; the right panel shows particles placed on ellipsoids. In both cases, a small random displacement is applied, since it leads to a better initial condition for the APM iteration, as already noted in \cite{diener22a}. The star belongs to a 1.9 $M_\odot$ equal-mass, $47.5$ km separation, MS1b, irrotational BNS produced with \lorene. The blue points have $|z|<\,\sim\!\! 0.16$ km and trace the surface of the star (for more details, see \autoref{app:ovals}). The particles placed on surface-conforming ovals model very accurately the shape of the star and produce a smooth surface, thus providing a more accurate initial condition for the APM.}
    \label{fig:ovals1919}
\end{figure*}

Until recently we have prepared the initial condition for the APM on each star by placing particles on spherical surfaces with radii in the interval $(0,R)$, with $R$ being the larger radius of the star, see \cite[Section~2.2.2]{diener22a}. This algorithm places close-to-equal-mass particles on each spherical surface, taking into account the mass of the spherical shell bounded by a spherical surface and the next. Therefore, some information on the density profile of the star is considered already at the level of the initial condition for the APM iteration.
We have improved our algorithm by placing the initial particles 
on ovals that conform to the (scaled) surface of the star, otherwise we follow
the same steps as described in \cite[Appendix~B.1]{diener22a}, more details can be found in Appendix~\ref{app:ovals}. A comparison between particle distributions, produced using ovals and ellipsoids, is shown in \autoref{fig:ovals1919} for a star whose geometry deviates significantly from spherical.
The placement of particles on ovals improves the accuracy of the particle model of the outer layers of the star. This is because the APM starts with initial conditions having a smoother particle distribution on the outer layers, compared to the distributions obtained using spheres or ellipsoids. The latter include cuts on the outer layers when the spheres or ellipsoids cut through the surface of the star. This improvement is even more important for rotationally flattened stars.

\FloatBarrier

\subsection{The boundary particles used during the APM}
\label{subsec:boundary}

\begin{figure*}[t]
    \centering
    \includegraphics[width=\linewidth]{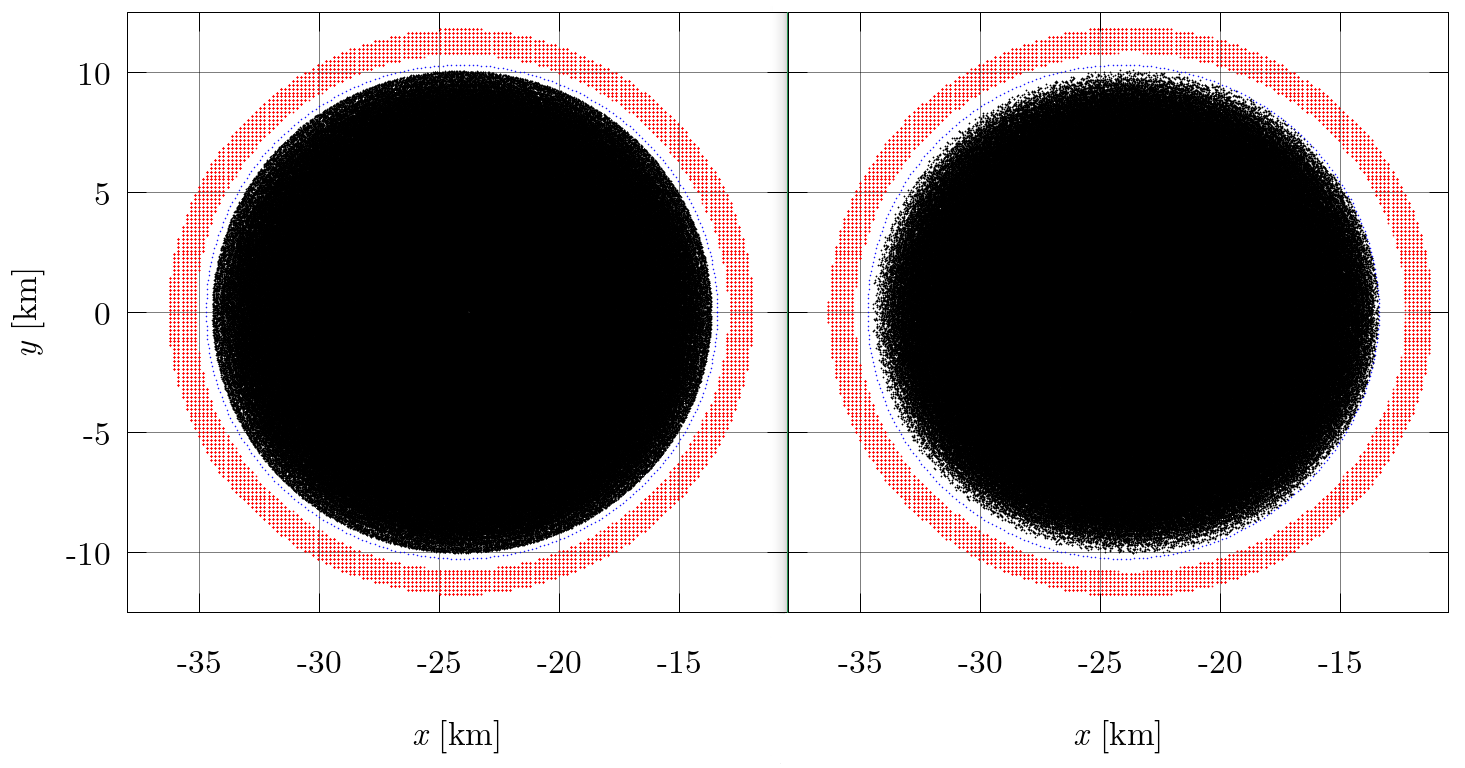} \\
    \includegraphics[width=\linewidth]{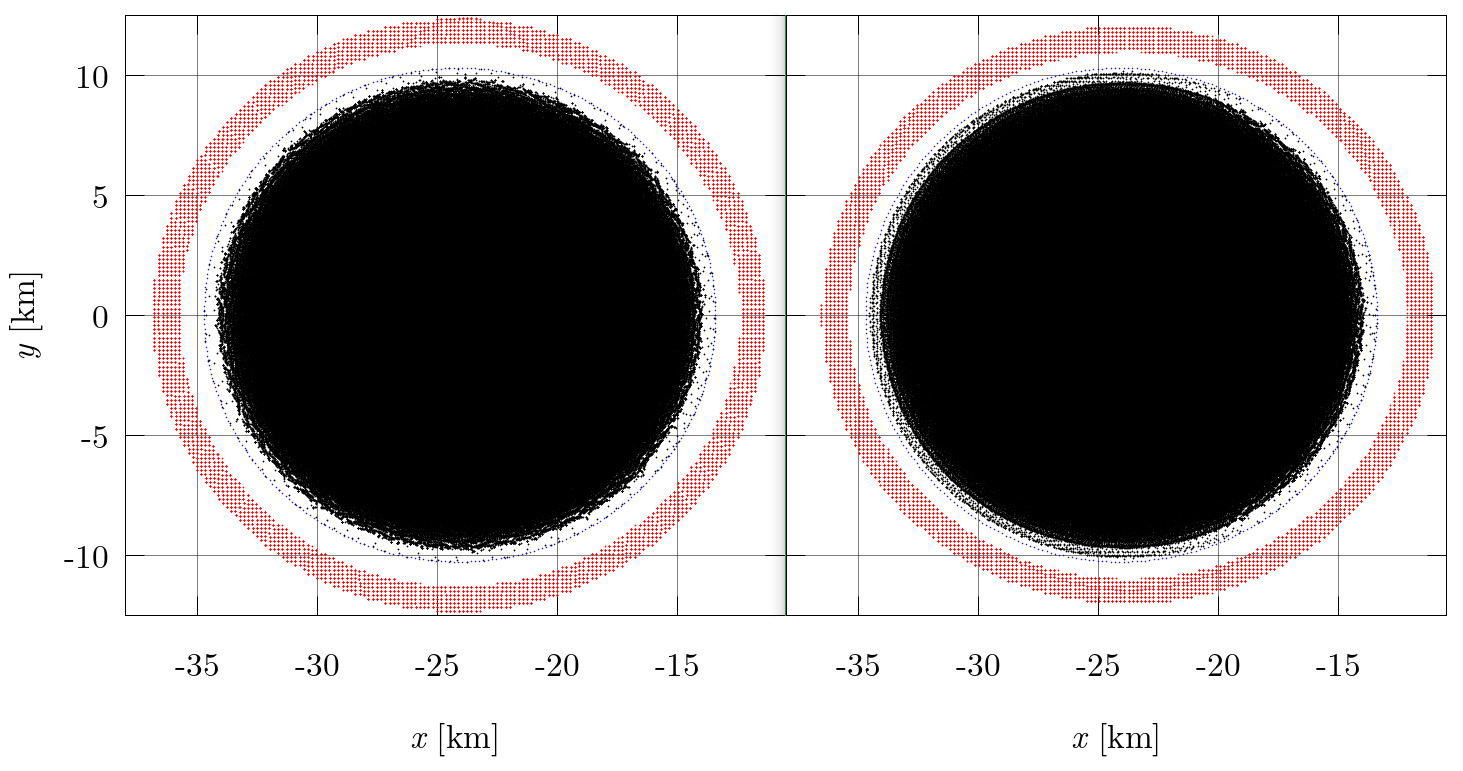}
    \caption{Projections of particle distributions with $2.5\times 10^6$ particles modeling the same neutron star. Real particles are in black, boundary particles with $|z|<\,\sim\!\! 0.74$ km are in red, and the points lying on the surface of the star with $|z|<\,\sim\! 0.16$ km are in blue. The star belongs to a $1.9 M_\odot$ equal-mass, $47.5$ km separation, MS1b, irrotational BNS produced with \lorene. {\bf(Top-left)} The real and boundary particles are placed on surface-conforming ovals, see Section~\ref{subsec:ovalsurfaces} and Section~\ref{subsec:boundary}. {\bf (Top-right)} The real and boundary particles are placed on ellipsoids. {\bf (Bottom-left)} Particles placed with our latest APM algorithm, with the particles in the top-left panel as initial condition. {\bf (Bottom-right)} Particles placed with our older APM algorithm, with the particles in the top-right panel as initial condition. See main text for details. The particles in the bottom-left panel better model the outer layers and the geometry of the star, compared to those in the bottom-right panel.}
    \label{fig:apm1919}
\end{figure*}

The APM iteration uses ``ghost" or ``boundary" particles outside the star that prevent the particles modelling the star from being pushed outside the stellar surface, see
 \cite[Section~2.2.2]{diener22a} and \cite[Appendix~B.2]{diener22a}.
At each step of the iteration, the boundary particles are assigned an artificial pressure that increases linearly with their distance from the center of the star. 
The boundary particles closest to the star are assigned an artificial pressure equal to $3\pi_\mathrm{max}$, those farthest away a value of $30\pi_\mathrm{max}$ and for those in between, the artificial pressure varies linearly between these bounds, where
 $\pi_\mathrm{max}$ be the maximum artificial pressure of the real particles. For these bounds we empirically found the best APM results.
 This artificial pressure gradient makes the real particles feel a stronger repulsive force, the 
 closer they approach the stellar surface. Previously \citep{diener22a}, we placed the boundary particles  on a lattice, between two ellipsoidal surfaces, now we place them on a lattice between two surface-conforming ovals instead. This, analogously to the initial placement of real particles on surface-conforming ovals described in Section~\ref{subsec:ovalsurfaces}, makes it easier for the APM iteration to model the outer layers and the overall geometry of the star.\\
In \cite[Appendix~B.2]{diener22a}, the parameter $\delta$ was introduced, which is the distance between the surface of the star and the boundary particle closest to it, along the direction of the star's largest radius. The modeling of the outer layers turns out to somewhat depend on the value of $\delta$, and it would be desirable to remove this dependence.  If $\delta$ is too small, the real particles cannot approach the surface of the star; if $\delta$ is too large,  the particle distribution on the outer layers can become non-smooth, leaving a few isolated particles
outside the otherwise smooth stellar surface.
In order to  reduce the dependence on $\delta$, we  set a small value of $\delta$ initially, and then let the boundary particles move outwards (effectively increasing $\delta$) very slowly until the condition
$|r_\mathrm{av} - R| < h_\mathrm{av}/3$ is met, where
$r_\mathrm{av}$ and $h_\mathrm{av}$ are the average values of radius and smoothing lengths of  the particles in the outer layers and $R$ is largest radius of the star.
 The particles modeling the outer layers are defined as those having a radial coordinate (measured from the center of the star) larger than $99.5\%$  
 of $R$. In case these should be less than 10 particles, this fraction is reduced in steps of 
 $0.5\%$ until this number is exceeded.
This algorithm starts out producing a smooth particle distribution on the outer layers since $\delta$ is initially small, and preserves the smoothness when it gently allows the real particles to move towards the surface. In addition, it reduces the dependence on the initial value of $\delta$ since the latter is increased during the iteration. \\
A comparison between a particle distribution obtained using the latest methods described in this section and in Section~\ref{subsec:ovalsurfaces}, and a particle distribution obtained with older methods, is shown in Figure~\ref{fig:apm1919}. Note how the model of the geometry of the star and its outer layers have improved with the new methods.

\subsection{The ADM linear momentum for the initial data}
\label{subsec:admlinmom}

When considering  ID produced with \lorene and \fuka, we can simplify the SPH estimate of the ADM momentum of the fluid, Equation~\eqref{eq:admmomesph}. The ID solvers assume asymptotic flatness, conformal flatness and maximal slicing on the initial spacelike hypersurface \cite{Gourgoulhon_2001,Papenfort_2021}. In coorbiting coordinates of Cartesian type, the conformal flatness condition can be written as \cite[Section~IV.A]{Gourgoulhon_2001},\cite[Equation~(6)]{Papenfort_2021}:
\begin{align}
\label{eq:confflat}
	\gamma_{ij}=A^2\delta_{ij},
\end{align}
with $A$ being the conformal factor and $\delta_{ij}=\mathrm{diag}(1,1,1)$ the Euclidean metric. Hence, a global, orthogonal, non-orthonormal frame $e^i_{(j)}=\delta^i{}_j$ exists on the initial spacelike hypersurface. This frame also becomes orthonormal, hence Cartesian, at spatial infinity where $A\rightarrow 1$ due to the asymptotic flatness condition. Therefore, we can set $\xi^i=e^i_{(j)}$ in Equation~\eqref{eq:admmomcons} to compute the $j^{\rm th}$ Cartesian component of the ADM linear momentum. Doing so, and imposing the maximal slicing condition, the part of the ADM linear momentum determined by the spacetime---i.e., the second term in the squared parenthesis in Equation~\eqref{eq:admmomcons}---is zero. We show this explicitly in \autoref{app:admmomlorene}.\\
Hence, for the  ID the total ADM linear momentum is equal to the ADM momentum of the fluid, which can be estimated with \eqref{eq:admmomesph} for the \lorene and \fuka  ID. In addition, we can compare this estimate with the one obtained using \eqref{eq:admmom}. It is possible to compute the linear ADM momentum within \lorene as a surface integral at infinity.\footnote{We added this feature to our fork of \lorene, since we could not find a function that does it in the original fork. However, a function that computes the Bowen--York angular momentum as described in \cite[Section~IIID]{Gourgoulhon_2001}, was included in the original fork \cite{binaire-angumom}, and we used it as a template.} \lorene can easily handle this computation thanks to its compactified coordinates. \fuka also provides an estimate of the ADM momentum. Hence, for each  ID, we have two independent estimates of the ADM linear momentum: one computed by the ID solver as a surface integral using \eqref{eq:admmom}, and the other computed by \spi as an SPH estimate of a volume integral using \eqref{eq:admmomesph}. For the irrotational, equal-mass $1.3M_\odot$ \lorene ID with 2 million SPH particles, see  Section~\eqref{sec:stable_remnant}, the two estimates are:
\begin{subequations}
\begin{alignat}{5}
\label{eq:admmom-comparison}
    \lorene :       &\quad P^\mathrm{ADM}_i    &&\simeq \big( < 10^{-15}&, \, 6.336\times 10^{-11}&, \, < 10^{-15}\big), \\
    \spi :   &\quad P^\mathrm{ADM}_i    &&\simeq \big(-1.393\times 10^{-7}&, \, 6.318\times 10^{-11}&, \, < 10^{-15}\big).
\end{alignat}
\end{subequations}
Only the $x$ component is affected by a substantial error after the particles are placed, but it stays very small nonetheless. We obtain similar results for the other equal-mass, non-spinning systems we consider in Section~\ref{sec:BH_formation}. For the equal-mass $1.3M_\odot$ \fuka ID used in the simulation described in Section~\ref{sec:single_Spin}, where  one star is spinning with $\chi\simeq 0.5$, the estimates are:
\begin{subequations}
\begin{alignat}{8}
\label{eq:admmom-comparisonf}
    \fuka :       &\quad P^\mathrm{ADM}_i    &&\simeq \big(- 2.804\times 10^{-13}&,& \,\, 1.683\times 10^{-10}&,& \, < 10^{-15}\big), \\
    \spi : &\quad P^\mathrm{ADM}_i    &&\simeq \big( \hphantom{lin}6.030\times 10^{-6}&,& \,\, 2.809\times 10^{-4}&,& \, < 10^{-15}\big).
\end{alignat}
\end{subequations}
It makes sense that the SPH estimate is much better for the non-spinning systems since the particles modeling the second star are placed mirroring those modeling the first star, with respect to the $yz$ plane \cite[Section~2.2]{diener22a}. Triggered by questions from the referee, we realized that it
would be even better to reflect the particle
positions of star one through the origin to set up 
star two. In this case particles would be placed 
with complete symmetry for all components of the 
coordinates (including $x$) and the helical 
symmetry present in the initial data would 
guarantee that the velocity of a particle in 
star two is exactly opposite the corresponding
particle in star one. There would then be exact 
cancellation (to roundoff error) of their
contribution to the momentum, allowing us to
reduce the small initial value for the 
$x$-component of the ADM momentum for irrotational, 
equal mass systems. This procedure will be used
in future particle setups. In the system with a spinning star, however, neither mirror nor reflection 
symmetry can be enforced, and the terms in the sum 
\eqref{eq:admmomesph} do not compensate to the 
same degree.

%% file: sphincs_id-appendices.tex
\section{Surface-conforming ovals}
\label{app:ovals}

In order to use the surface-conforming ovals as discussed in Sec.~\ref{subsec:ovalsurfaces}, we first need to find the surface of the star. We do this in \spi in the following way.
\begin{enumerate}
	\item Choose a threshold density that marks the end of the star. Currently, this value is $\rho_\mathrm{thres}=10^{-12}\,\mbox{code units}\simeq 6.2\times 10^{5}\,\mathrm{g/cm^3}$ (see the beginning of Sec.~\ref{sec:hydro} for the definition of our code units). Also, choose a tolerance value $\mathsf{tol}= 10^{-6}$ (code units), the choices of these values are motivated in step 3.
	\item Select a direction $(\theta,\phi)$ centered at the center of the star, and choose two points lying on it, one inside and one outside the star. Read the density provided by the ID solver at these points. 
	\item Apply the bisection algorithm around $\rho_\mathrm{thres}$ until the distance between the two points is equal to or lower than $\mathsf{tol}$. Then, the innermost of the two points is declared to be the radius $r$ along the considered $(\theta,\phi)$ direction. Arguably, the values of $\rho_\mathrm{thres}$ and $\mathsf{tol}$ are somewhat arbitrary. We determined them by comparing the resulting values for the 4 radii along the positive and negative $x$ direction, and along the $y$ and $z$ direction, with the values provided by \lorene. 
	\item Repeat steps 2 and 3 for the desired number of directions, to find a set of points $\{\theta,\phi,r\}$ that lie on the surface of the star. Currently, we sample the surface every degree in both $\theta$ and $\phi$; we thus have $180\times 360$ points.
\end{enumerate}
Once the sample $\{\theta,\phi,r\}$ lying on the surface is found, the value of $r(\theta,\phi)$ for any $(\theta,\phi)\in [0,\pi)\times [0,2\pi)$ is found with bilinear interpolation.

\section{The SPH estimate of the ADM linear momentum of the fluid}
\label{app:admmomsph}
Here we explicitly compute the ADM linear momentum of the fluid in terms of the SPH canonical momentum, as mentioned in Sec.~\ref{subsec:admlinmom}.
The integral in \eqref{eq:admmom} can be written as an integral over the spacelike hypersurface $\Sigma$ using Gauss' theorem,
\begin{align}
\label{eq:admmomgauss}
	P_\xi^{\rm ADM}&\coloneqq \dfrac{1}{8\pi}\int_{\partial\sigma}\dd S_j\left(K^j{}_i - \delta^j{}_i \,K\right)\xi^i \nonumber \\
    &= \dfrac{1}{8\pi}\int_{\Sigma}\dd ^3x\sqrt{\gamma}\left[\xi^iD_j\left(K^j{}_i - \delta^j{}_i \,K\right)+\left(K^j{}_i - \delta^j{}_i \,K\right)D_j\xi^i\right], 
\end{align}
with $\gamma$ determinant of the spatial metric $\gamma_{ij}$, and $D_j$ covariant derivative compatible with $\gamma_{ij}$. The second term in the square parenthesis can be rewritten as
\begin{align}
\label{eq:liemetric}
    \left(K^j{}_i - \delta^j{}_i \,K\right)D_j\xi^i 
    &= \left(K^{ij} - \gamma^{ij} \,K\right)D_j\xi_i \nonumber \\
    &= \left(K^{ij} - \gamma^{ij} \,K\right)\dfrac{D_j\xi_i+D_i\xi_j}{2}
    = \left(K^{ij} - \gamma^{ij} \,K\right)\dfrac{\mathscr{L}_\xi\gamma_{ij}}{2},
\end{align}
where the first equality uses the compatibility between $\gamma_{ij}$ and $D_j$, the second equality uses the symmetry of $K^{ij}$ and $\gamma^{ij}$ in their indices, and the last equality uses the definition of the Lie derivative $\mathscr{L}_\xi\gamma_{ij}$. The first term in the square parenthesis in \eqref{eq:admmomgauss} can be rewritten using the momentum constraint \cite[eq.(2.126)]{baumgarte10},
\begin{align}
\label{eq:momconstraint}
    D_j\left(K^j{}_i - \delta^j{}_i \,K\right)=
    8\pi \eulS_i,
\end{align}
where $\eulS_i$ is the spatial part of the momentum density measured by the Eulerian observer $\eulS_\rho\coloneqq -n^\mu\,T_{\mu\nu}\,\gamma^\nu{}_\rho$, with $T_{\mu\nu}$ being the stress--energy tensor, $n^\mu=(1, -\beta^i)/\alpha$ vector normal to $\Sigma$ (and 4-velocity of the Eulerian observer), $\alpha$ lapse function, $\beta^i$ shift vector, and $\gamma^\mu{}_\nu= \delta^\mu{}_\nu + n^\mu n_\nu$ projector onto $\Sigma$. Hence, the ADM linear momentum in \eqref{eq:admmomgauss} can be rewritten as
\begin{align}
\label{eq:admmomcons2}
    P_\xi^{\rm ADM}&= \int_{\Sigma}\dd ^3x\sqrt{\gamma}
    \left[\xi^i\eulS_i + \dfrac{1}{16\pi} \left(K^{ij} - \gamma^{ij} \,K\right)\mathscr{L}_\xi\gamma_{ij}\right].
\end{align}
Now we use the definition of $\eulS_\mu$ to write
\begin{align}
\label{eq:momdens}
    \eulS_\mu\coloneqq-n^\rho\,T_{\rho\nu}\,\gamma^\nu{}_\mu 
    &=-n^\rho T_{\rho\mu} - \left(n^\rho T_{\rho\nu}n^\nu\right)n_\mu.
\end{align}
The spatial part of $\eulS_\mu$ is (see Eq.~\eqref{eq:BSSN_Si})
\begin{align}
\label{eq:momdens-spatial}
    \eulS_i 
    &=-n^\nu T_{\nu i} - \left(n^\nu T_{\nu\rho}n^\rho\right)n_i = -n^\nu T_{\nu i}
    = \dfrac{\beta^j T_{ji} - T_{0i}}{\alpha}
\end{align}
where we used $n_\mu=(-\alpha, 0,0,0)$ and the expression for the components of $n^\mu$, given above.\\
The SPH canonical momentum per baryon, for a perfect fluid, is defined as (see, e.g., \cite[eq.~198]{rosswog09b})
\begin{align}
\label{eq:sphmom}
    \sphS_i\coloneqq \Theta \left( 1+u+\dfrac{P}{n} \right)g_{i\mu}v^\mu,
\end{align}
where $v^\mu=U^\mu/U^0$ is the fluid velocity in the SPH computing frame, $U^\mu=\dd x^\mu/\dd\tau$ is the 4-velocity of the fluid, $\tau$ is the proper time of the fluid, $\Theta$ is the generalized Lorentz factor $\Theta\coloneqq \left(-g_{\mu\nu}v^\mu v^\nu\right)^{-1/2}=U^0$, with $g_{\mu\nu}$ spacetime metric constructed from $\gamma_{ij}$, $\alpha$ and $\beta^i$; $u$ is the specific internal energy, $P$ the pressure, and $n$ the baryon density of the fluid. The last three hydrodynamical quantities are measured in the local rest frame of the fluid. It holds:
\begin{align}
\label{eq:vcov}
	U_i&= g_{i\mu}U^\mu= g_{i0}U^0 + g_{ij}U^j= U^0\beta_i + \gamma_{ij}U^j \Longrightarrow \gamma_{ij}U^j= U_i - U^0\beta_i, \nonumber \\
	g_{i\mu}v^\mu&= g_{i\mu}\dfrac{U^\mu}{U^0}=\dfrac{g_{i0}U^0 + g_{ij}U^j}{U^0}= g_{i0} + \dfrac{\gamma_{ij}U^j}{U^0}= \beta_i + \dfrac{U_i - U^0\beta_i}{U^0}= \dfrac{U_i}{U^0}.
\end{align}
Using \eqref{eq:vcov} in \eqref{eq:sphmom},
\begin{align}
\label{eq:sphmom2}
    \sphS_i= \left( 1+u+\dfrac{P}{n} \right)U_i.
\end{align}
Since the SPH canonical momentum is a momentum per baryon, we can try to build it from the stress--energy tensor in the following way:
\begin{align}
\label{eq:mombar}
    q_i\coloneqq \dfrac{T_{0i}}{N},
\end{align}
where $N\coloneqq \sqrt{-g}\,\Theta\, n$ is the baryon number density in the computing frame, with $g$ determinant of $g_{\mu\nu}$. Dimensionally, $q_i$ is a momentum per baryon. The expression \eqref{eq:sphmom} holds for a perfect fluid, hence we now specialize to the stress--energy tensor of a perfect fluid,
\begin{align}
\label{eq:mombar2}
    q_i= \dfrac{(e+P)U_0U_i +P\,g_{0i}}{N},
\end{align}
with $e=n(1+u)$ energy density of the fluid measured in its local rest frame. We use
\begin{align}
\label{eq:u0}
    U_0= g_{0\mu}U^\mu = g_{00}U^0 + g_{0i}U^i = g_{00}\Theta + g_{0i}\Theta v^i
       = \Theta\left(-\alpha^2 + \beta^j\beta_j + \beta_j v^j\right),
\end{align}
in expanding \eqref{eq:mombar2},
\begin{align}
\label{eq:mombar3}
    q_i &= \dfrac{e+P}{\sqrt{-g}\Theta \, n}\Theta\left(-\alpha^2 + \beta^j\beta_j + \beta_j v^j\right)U_i + \dfrac{P}{\sqrt{-g}\Theta\, n}\beta_i.
\end{align}
We substitute $e=n(1+u)$ into \eqref{eq:mombar3},
\begin{align}
\label{eq:mombar4}
    q_i    &= \dfrac{1}{\sqrt{-g}}\left[\left(1+u+\dfrac{P}{n}\right)U_i\right]\left(-\alpha^2 + \beta^j\beta_j + \beta_j v^j\right) + \dfrac{P}{\sqrt{-g}\Theta\, n}\beta_i 
    \nonumber \\                            
    &=\dfrac{1}{\sqrt{-g}}\left[\sphS_i
    \left(-\alpha^2 + \beta^j\beta_j + \beta_j v^j\right) + \dfrac{P}{\Theta\, n}\beta_i\right].
\end{align}
Using \eqref{eq:mombar}, the expression in \eqref{eq:mombar4} becomes
\begin{align}
\label{eq:mombar5}
    T_{0i}
    =\dfrac{N}{\sqrt{-g}}\left[\sphS_i
    \left(-\alpha^2 + \beta^j\beta_j + \beta_j v^j\right) + \dfrac{P}{\Theta\, n}\beta_i\right] 
    =\Theta\, n\left[\sphS_i
    \left(-\alpha^2 + \beta^j\beta_j + \beta_j v^j\right) + \dfrac{P}{\Theta\, n}\beta_i\right],
\end{align}
that is, we related the SPH momentum per baryon $\sphS_i$ with the $T_{0i}$ components of the perfect fluid stress--energy tensor. Note that, if the expression for the SPH momentum per baryon was known in general, the same computation could in principle be generalized to any type of fluid.\\
We can now insert \eqref{eq:mombar5} into \eqref{eq:momdens-spatial}, and obtain the relation between the Eulerian momentum density $\eulS_i$ and the SPH momentum per baryon $\sphS_i$,
\begin{align}
\label{eq:momrelationstep}
    \eulS_i 
    &= \dfrac{\beta^j T_{ji} - T_{0i}}{\alpha} \nonumber \\
    &= \dfrac{\beta^j T_{ji}}{\alpha}
       - \dfrac{\Theta\, n}{\alpha}\left[\sphS_i
    \left(-\alpha^2 + \beta^j\beta_j + \beta_j v^j\right) + \dfrac{P}{\Theta\, n}\beta_i\right] \nonumber \\ 
    &= \dfrac{\beta^j}{\alpha}\left[n\left(1+u+\dfrac{P}{n}\right)U_i U_j + P\, \gamma_{ij}\right]
       - \dfrac{\Theta\, n}{\alpha}\left[\sphS_i
    \left(-\alpha^2 + \beta^j\beta_j + \beta_j v^j\right) + \dfrac{P}{\Theta\, n}\beta_i\right] \nonumber \\ 
    &= \dfrac{\beta^j U_j \,n}{\alpha}\sphS_i  - \dfrac{\Theta \,n}{\alpha}\sphS_i \,\beta_j v^j + \dfrac{P\, \beta_i}{\alpha} - \dfrac{P\, \beta_i}{\alpha}
       - \dfrac{\Theta\, n}{\alpha}
    \left(-\alpha^2 + \beta^j\beta_j\right)\sphS_i
     \nonumber \\ 
    &= \dfrac{\sphS_i\,n}{\alpha}
    \left(\beta^jU_j + \Theta\alpha^2 - \Theta\beta^j\beta_j - \Theta\beta_j v^j\right).
\end{align}
Using \eqref{eq:vcov}, we rewrite
\begin{align}
\label{eq:contraction}
	\beta^j U_j= \beta^j \Theta g_{j\mu}v^\mu= \Theta \left(\beta^j g_{j0}v^0 + \beta^j g_{jk}v^k\right)= \Theta \left(\beta^j \beta_j + \beta_k v^k\right),
\end{align}
where we used $v^0=U^0/U^0=1$. Substituting \eqref{eq:contraction} into \eqref{eq:momrelationstep},
\begin{align}
\label{eq:momrelation}
	\eulS_i= \dfrac{\sphS_i\,\Theta\, n}{\alpha}\left(\beta^j \beta_j + \beta_k v^k + \alpha^2 - \beta^j\beta_j - \beta_j v^j \right)= \Theta\, n\,\alpha\, \sphS_i.
\end{align}
Using \eqref{eq:momrelation}, the expre/ssion for the ADM momentum determined by the fluid becomes [see \eqref{eq:admmomcons}]
\begin{align}
\label{eq:admmomfinal}
    P_\xi^{\rm ADM,fluid}&= \int_{\Sigma}\dd ^3x\sqrt{\gamma}\,\xi^i S_i = \int_{\Sigma}\dd ^3x\left(\sqrt{\gamma}\,\Theta\, n\alpha \right)\xi^i \sphS_i = \int_{\Sigma}\dd ^3x \,N \,\xi^i \sphS_i.
\end{align}
where we used $\sqrt{-g}= \alpha\sqrt{\gamma}$ and $N\coloneqq \sqrt{-g}\,\Theta\, n$.\\
The integral in \eqref{eq:admmomfinal} is over the entire spacelike hypersurface $\Sigma$, but it has support only where $\sphS_i$ (or $N$) is nonzero, that is, where the fluid is. Therefore, we can approximate the integral with an SPH summation over the particles, to obtain an estimate of the ADM linear momentum of the fluid,
\begin{align}
\label{eq:admmomsph}
    P_\xi^{\rm ADM,fluid}&= \int_{\Sigma}\dd ^3x \,N \,\xi^i \sphS_i 
    \simeq \sum_a \nu_a \left(\xi^i \sphS_i \right)_a,
\end{align}
with the index $a$ running over all the particles, $\nu_a$ being the baryon number of particle $a$.

\section{Computation of the ADM linear momentum for the ID}
\label{app:admmomlorene}
%
Both \lorene and \fuka assume a spacetime free of
gravitational waves which implies that the whole 
momentum should be carried by the fluid alone. We briefly show this here explicitly.\\
As discussed in \eqref{subsec:admlinmom}, on the initial spacelike hypersurface and with the assumptions of asymptotic flatness, conformal flatness and maximal slicing, used by \lorene and \fuka to solve for the ID, $e^i_{(j)}=\delta^i{}_j$ is an orthogonal frame which becomes orthonormal, hence Cartesian, at spatial infinity. We can then set $\xi^i=e^i_{(j)}$ in \eqref{eq:admmomcons}, and obtain a formula for the Cartesian components of the ADM linear momentum at $t=0$:
\begin{align}
\label{eq:admmomid}
    P_j^{\rm ADM}&= \int_{\Sigma}\dd ^3x\sqrt{\gamma}
    \left[e^i_{(j)}\eulS_i + \dfrac{1}{16\pi} \left(K^{mi} - \gamma^{mi} \,K\right)\mathscr{L}_{e_{(j)}}\gamma_{mi}\right].
\end{align}
We now use \eqref{eq:confflat} to note that
\begin{align}
\label{eq:LieCart}
	\mathscr{L}_{e_{(j)}}\gamma_{mi}	&= \mathscr{L}_{e_{(j)}}(A^2\delta_{mi})=\delta_{mi}\,\mathscr{L}_{e_{(j)}}A^2 + A^2\,\mathscr{L}_{e_{(j)}}\delta_{mi} \nonumber \\
	&=\delta_{mi}e^k_{(j)}\partial_k A^2 + A^2\big(e^k_{(j)}\partial_k\delta_{mi} + \delta_{ik}\partial_m e^k_{(j)} + \delta_{mk}\partial_i e^k_{(j)}\big)= \delta_{mi}e^k_{(j)}\partial_k A^2,
\end{align}
were the last equality follows from $\delta_{mi}$ and $e^k_{(j)}$ having constant components. Inserting \eqref{eq:LieCart} into \eqref{eq:admmomid},
\begin{align}
\label{eq:admmomidcart}
    P_j^{\rm ADM}&= \int_{\Sigma}\dd ^3x\sqrt{\gamma}
    \left[e^i_{(j)}\eulS_i + \dfrac{1}{16\pi} \left(K^{mi} - \gamma^{mi} \,K\right)\delta_{mi}e^k_{(j)}\partial_k A^2\right].
\end{align}
Finally, using the maximal slicing condition $K=K^i{}_i=K^{mi}\gamma_{mi}=A^2 K^{mi}\delta_{mi}=0$, which implies $K^{mi}\delta_{mi}=0$ because $A\neq 0$, 
\begin{align}
\label{eq:admmomidcart-final}
    P_j^{\rm ADM}&= \int_{\Sigma}\dd ^3x\sqrt{\gamma}\,e^i_{(j)}\eulS_i=\int_{\mathrm{stars}}\dd ^3x\sqrt{\gamma}\,\eulS_j.
\end{align}
Hence, for the \lorene and \fuka binary neutron star ID, it suffices to integrate \eqref{eq:admmomidcart-final} over a compact domain (the stars), to obtain the total linear ADM momentum.